\documentclass[10pt,journal,compsoc]{IEEEtran}

\newcommand{\tool}{{Pelican}\xspace}

\usepackage{multirow}
\usepackage{booktabs}
\usepackage{pdfpages}
\usepackage{graphicx,dblfloatfix}
\usepackage[linesnumbered,boxed]{algorithm2e}
\usepackage{tikz}
\usepackage{calc}
\usepackage{amssymb}
\usepackage{pifont}
\newcommand{\cmark}{\ding{51}}%
\newcommand{\xmark}{\ding{55}}%
\usepackage{amsmath}

\usepackage{enumitem}
\usepackage{xcolor}
\usepackage{color}
\usepackage{multicol}
\usepackage{wrapfig}
\usepackage{hyperref}
\usepackage{subcaption}
\usepackage{graphicx}
\definecolor{mygreen}{rgb}{0,0.6,0}
\definecolor{mymauve}{rgb}{0.58,0,0.82}
\usepackage[figuresright]{rotating}
\usepackage{colortbl}
\definecolor{mygray}{gray}{.85}

\usepackage{balance}
\usepackage{verbatim}

\usepackage{listings}

\begin{document}

\title{Advanced Evasion Attacks and Mitigations on Practical ML-Based Phishing Website Classifiers}

\author{\IEEEauthorblockN{Yusi Lei,
		Sen Chen,
		Lingling Fan,
		Fu Song, and
		Yang Liu}
\IEEEcompsocitemizethanks{
	\IEEEcompsocthanksitem
	Yusi Lei and Fu Song (Corresponding Author) are with School of Information Science and Technology, 
	ShanghaiTech University, China.
	Emails: yusi.lei.ecnu@gmail.com and songfu@shanghaitech.edu.cn
	\IEEEcompsocthanksitem
	Sen Chen, Linglnig Fan, and Yang Liu are with the School of Computer Science and Engineering, Nanyang Technological University, Singapore.
	Emails: \{chensen, llfan, yangliu\}@ntu.edu.sg
	\IEEEcompsocthanksitem
	Sen Chen and Yusi Lei are the co-first authors.
}}

\IEEEtitleabstractindextext{
\begin{abstract}
	Machine learning (ML) based approaches have been the mainstream solution for anti-phishing  detection.
	When they are deployed on the client-side, ML-based classifiers are vulnerable to evasion attacks, as shown by recent attacks.
	However, such potential threats have received relatively little attention because existing attacks destruct the functionalities or appearance of webpages and are conducted in the white-box scenario,
	making it less practical. Consequently, it becomes imperative to understand whether it is possible to launch evasion attacks with limited
knowledge of the classifier, while preserving the functionalities and appearance.
	
	In this work, we show that even in the grey-, and black-box scenarios, evasion attacks are not only effective on practical ML-based classifiers, but
	can also be efficiently launched without destructing the functionalities and appearance.
	For this purpose, we propose three mutation-based attacks, differing in the knowledge of the target classifier,
	addressing a key technical challenge: automatically crafting an adversarial sample from a known phishing website in a way that can mislead classifiers.
	To launch attacks in the white- and grey-box scenarios,
	we also propose a sample-based collision attack to gain the knowledge of the target classifier.
	We demonstrate the effectiveness and efficiency of our evasion attacks on the state-of-the-art, Google's phishing page filter, achieved 100\% attack success rate in less than one second per website.
	Moreover, the transferability attack on BitDefender's industrial phishing page classifier, TrafficLight, achieved up to 81.25\% attack success rate.
	We further propose a similarity-based method to mitigate such evasion attacks, \tool, which compares the similarity of an unknown website with recently detected phishing websites. We demonstrate that \tool can effectively detect evasion attacks, hence could be integrated into ML-based classifiers. We also highlight two strategies of classification rule selection to enhance the robustness of classifiers.
	Our findings contribute to design more robust phishing website classifiers in practice.
\end{abstract}}

\maketitle

\IEEEdisplaynontitleabstractindextext
\IEEEpeerreviewmaketitle

\IEEEraisesectionheading{\section{Introduction}\label{Sec:Intr}}
\IEEEPARstart{W}{eb} phishing is a major cyber threat and have reached a record volume~\cite{PXQ0VW19,hong2012state}.
It has been widely used to steal personal information such as login credentials.
In addition, people nowadays prefer online shopping, which aggravates the harms of phishing attacks in practice.
Consequently, phishing attackers are able to gain a (vast) amount of money via phishing websites~\cite{Mathews17}.
To detect phishing
attacks, various anti-phishing solutions have been proposed, such as
blacklists~\cite{oest2019phishfarm},
heuristic-~\cite{teraguchi2004client,zhang2007cantina},
similarity-~\cite{AfrozG11,medvet2008visual,fu2006detecting,rosiello2007layout},
and machine learning (ML)-based~\cite{pan2006anomaly,basnet2008detection} approaches.
However,  blacklist is ineffective in detecting 0-day attacks~\cite{oest2019phishfarm};
similarity- and heuristic-based approaches being capable of detecting 0-day attacks,
have limited scalability and accuracy~\cite{teraguchi2004client,zhang2007cantina}.
ML-based ones are not only scalable and accurate, but can also detect 0-day attacks,
hence have been widely studied~\cite{ma2009beyond,whittaker2010large,XHRC11,corona2017deltaphish,ubing2019phishing} and
deployed in industrial browsers (e.g., \emph{Chrome} and \emph{Edge}~\cite{edge20}).

Prior research shows that ML-based classifiers are vulnerable to evasion attacks, e.g., \cite{SL14,xu2016automatically,chen2018automated,hu2018black}.
Such attacks have been extensively studied in image recognition and malware detection,
but little has done in anti-phishing.
This is potentially due to the new challenge
in this domain:
unlike \textit{adversarial images} which only need to preserve the appearance and \textit{adversarial malware} which only need to preserve the functionality,
adversarial phishing websites have to preserve appearance and functionalities \emph{simultaneously}, to be more effective for web phishing~\cite{whittaker2010large}.
Although, a recent attack in this domain
has been exhibited in the white-box scenario~\cite{liang2016cracking},
it received relatively little attention, due to the following two limitations.
(1) Their attack might \emph{destruct} functionalities and appearance of webpages;
and (2) they assume that an attacker has
almost full knowledge of the target classifier, for which more effective obfuscation techniques~\cite{wroblewski2002general} could be used to protect classifiers.
Hence, it becomes imperative to understand whether it is possible to effectively launch evasion attacks on ML-based anti-phishing classifiers with limited
knowledge of the target classifier, while preserving the functionalities and appearance of the webpages.

In this work, we show that evasion attacks are not only effective on practical ML-based classifiers,
but can also be launched without destructing functionalities and appearance of phishing webpages, in all white-, grey-, and black-box scenarios.
For this purpose, we propose advanced evasion attacks, differing in
the knowledge (e.g., classification rules, features and weights) of a target classifier obtained by the adversary,
namely, white-box attack with full knowlegde, grey-box attack with partial knowledge, and black-box attack without any knowledge, of the ML-based classifier under attack.
Specifically, to effectively and efficiently craft adversarial samples from known phishing websites,
we leverage mutation techniques~\cite{jia2011analysis},  which are widely-used in fuzz testing~\cite{godefroid2008automated}.
In our setting,
we propose three mutation mechanisms, each of which iteratively mutates a sample from a known phishing website according to
the knowledge of the adversary until the target classifier is misled.
To launch attacks in white- and grey-box scenarios, it is necessary to gain partial or full knowledge of the classifier under attack.
Therefore, we propose a sample-based collision attack for inferring classification rules
by leveraging the data collected from
legitimate and phishing websites.
In the white-box attack setting, where (almost) full
knowledge of the target classifier is obtained, we leverage a greedy algorithm~\cite{hazewinkel2001greedy} to choose
which features to delete from or add into a sample in order to maximally decrease the decision score at each mutation step.
In the grey-box attack setting, where only partial knowledge of the target classifier is obtained,
we first iteratively delete all the features of known classification rules that can reduce the decision score,
and if necessary, we add classification rules that do not exist in original sample but can reduce the decision score. 
In the black-box setting, where no classifier information is exposed except that the adversary can query
the classifier and obtain the decision score, 
based on our insights of existing ML-based classifiers,
we first delete DOM nodes (i.e., nodes in HMTL Document Object Model) from the sample and then add DOM nodes into the sample, guided by the decision scores.

To evaluate our attacks,
we examined academic and industrial ML-based tools (e.g.,
CANTINA+~\cite{XHRC11}, Off-the-Hook~\cite{MAGSSA17}, $\delta$Phish~\cite{corona2017deltaphish}, the tool developed by Ubing et al.~\cite{ubing2019phishing}, Monarch~\cite{ThomasGMPS11}
and Google's phishing page filter (\texttt{GPPF})~\cite{whittaker2010large})
and industrial proprietary tools (Bitdefender {TrafficLight}~\cite{trafficlight}, Netcraft Anti-Phishing Extension~\cite{netcraft} and 360 Internet Protection~\cite{360protection}).
Except for \texttt{GPPF}, all the other tools either are not publicly accessible or cannot output decision score for each input which is required even in our black-box scenario.
Thus, we demonstrate our attacks
on the state-of-the-art ML-based classifier \texttt{GPPF}. We emphasize that many ML-based classifiers such as CANTINA+, Off-the-Hook and Monarch
use similar idea as \texttt{GPPF}, differ in features, architectures, and machine learning algorithms.
Therefore, our attacks could be used to these ML-based classifiers.
\texttt{GPPF} uses a logistic regression learning algorithm to train a classification model and is effective in phishing warnings~\cite{AF13}.
It has been deployed in both Chrome and Firefox~\cite{safebrowsing}, owning billions of users.
In terms of false positive rate and accuracy,
\texttt{GPPF} is also the best one among the 14 tools including similarity-, heuristic- and ML-based ones~\cite{MAGSSA17}.
To show that our black-box attack is generic and effective for other practical ML-based classifiers,
we launch transferability attacks on the industrial tool Bitdefender {TrafficLight}, which
owns the most users and stars in the add-ons of Firefox.
Furthermore, Bitdefender {TrafficLight} is proprietary without any publicly
available information about the internal design and implementations,
hence completely black-box.
We also tried to launch transferability attacks on other anti-phishing tools, but failed to get their classification results \emph{automatically}.
In summary, our attacks have 4 prominent advantages: 

\begin{itemize}
\item {\bf Effectiveness}. Our attacks achieve 100\% attack success rate on \texttt{GPPF} in all the white-, grey-, and black-box scenarios.
\item {\bf Efficiency}. 
Adversarial samples are crafted in less than one second per seed in average.
\item {{\bf Transferability}}. The transferability attack success rate on Bitdefender {TrafficLight} is up to 81.25\%.
\item {\bf Nondestructiveness}. All the crafted adversarial samples have the same functionalities and appearance as their original ones,
where appearance is measured by two common distortion criterions of images: mean-squared error (MSE)~\cite{wb09}
and mean-absolute error (MAE)~\cite{chai2014root}.
\end{itemize}

To mitigate the realistic threats posed by evasion attacks,
we propose a similarity-based method, named \tool, which compares the similarity of DOM tress between an unknown one and recently detected phishing websites.
This method neither modifies the target classifier nor relies on specific properties of the classifier,
so it could be used to protect a wide range of classifiers as a pre-filter.
We remark that \tool is used to efficiently detect evasion attacks rather than general phishing attacks as done by existing work (e.g., ~\cite{LiuDHF06,medvet2008visual,AfrozG11,fu2006detecting,rosiello2007layout}).
Furthermore, to mitigate collision attacks and enhance the robustness of classifiers,
we identify weaker classification rules and propose two strategies: \textit{removing single classification rules and subset classification rules from classifiers}.
We also suggest adding more \textit{appearance-related features} into these classification rules to increase the robustness of the classifier.
Our findings contribute to design more robust phishing website classifiers in practice.
We evaluate \tool on 915 adversarial samples (one sample per seed and scenario). 
\tool is able to detect all these samples.
We also evaluate classification rule selection strategies
using our black-box attacks and demonstrate their weakness,
but have no side-effect on checking original phishing/benign websites when subset rules or single rules are disabled. 

In this work, we make the following contributions.
\begin{itemize}
     \item	Novel evasion attacks on ML-based web phishing classifiers, which are effective, efficient, and nondestructive.

	\item A sample-based collision attack which can effectively infer more features/rules for grey- and white-box evasion attacks.
It is more effective and efficient than the recent collision attack technique~\cite{liang2016cracking}.
	
 	 \item Evaluation of our attacks which
 achieved 100\% attack success rate on the state-of-the-art classifier \texttt{GPPF} and achieved up to 81.25\%
 transferability attack success rate on Bitdefender {TrafficLight}, in the real world.

	\item A defense method, \tool, to mitigate  evasion attacks, and two strategies (i.e., removing single classification rules and subset classification rules) to
 enhance the robustness of the classifiers.

	\item A dataset~\cite{Pelican} including
	3,000 phishing websites, 15,000 legitimate website URLs and 2,566,834 phishing URLs to foster further research such as phishing website detection/defense and measurement.   
	
\end{itemize}

\medskip
To the best of our knowledge, this is the first study of evasion attacks on
practical ML-based phishing website classifiers in the grey- and black-box scenarios, the first defense work against evasion attacks in this domain
and the first evaluation of robustness of single classification rules and subset classification rules
under the adversarial environment.


\section{Background}\label{sec:background}


\subsection{Website and Webpage Structure} \label{sec:web_structure}
\begin{figure}[h]
	\centering\vspace{-3mm}
  \includegraphics[width=0.3\textwidth]{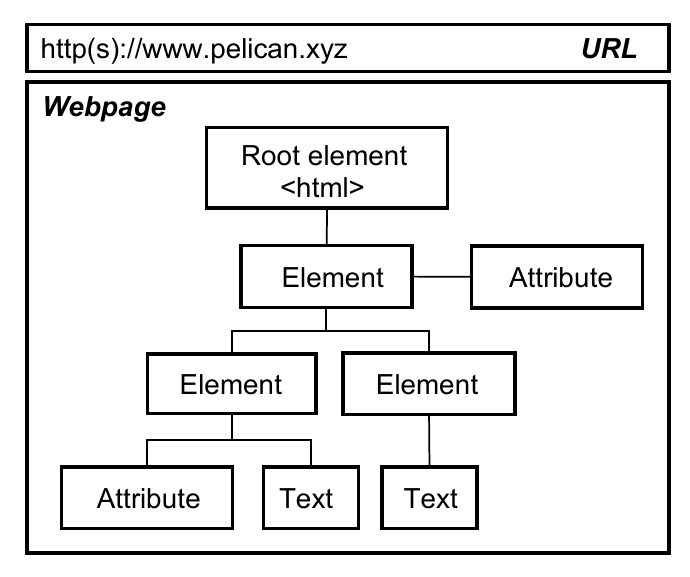}\vspace{-3mm}
 \caption{Structure of a webpage.}
	\label{fig:structure}\vspace{-1mm}
\end{figure}

A website consists of webpages for presenting information to users.
Webpages are accessed via their corresponding URLs.
As shown in Fig.~\ref{fig:structure}, a webpage has a basic structure
and {multiple} corresponding elements, representing in a tree hierarchy, called \emph{DOM tree}.
Specifically, a webpage has several kinds of nodes, called DOM nodes, including element (e.g., HTML element), text  (e.g., text inside the HTML element), attribute (e.g., the attributes of each HTML element), and comment codes (no effect on webpage structure).

The children of an element node are text and attribute nodes. 
In addition, scripts such as \texttt{JavaScript} code are also an important part of webpages
which can manipulate (e.g., \emph{modify}, \emph{delete}, and \emph{add}) the DOM tree.
In this work, we consider JavaScript only, other scripting languages can be handled similarly.
The appearance and functionalities of a webpage are determined by the combination of elements, texts, attributes, and scripts.

\subsection{Web Phishing}
Web phishing, a type of social engineering attacks~\cite{krombholz2015advanced, chen2019gui}, is often used to steal user information, such as login credentials and payment data~\cite{most}.
It occurs when an attacker masquerades the phishing website as a trusted entity, both of the webpages and URLs are very close to the original legitimate ones~\cite{kirda2006protecting}.
Typically, attackers first create a fake website and 
then distribute its URL to lure users to click the URL link, which redirects to a phishing website from untrusted servers to extract users' information~\cite{gupta2018defending} .
Attackers aim to make the phishing URLs and visual appearances of webpages as similar as possible to the targeted victim's.

Two traditional strategies are often adopted in web phishing: \textit{spoofed emails} and \textit{fake websites}~\cite{gupta2018defending, tang2019large}.
Typically, attackers first create a fake website and 
then send numerous spoofed emails to various users.
The spoofed emails lure users to click links in the email that redirects to a phishing website from untrusted servers to extract users' information.
Attackers aim to make the phishing URLs and visual appearances of webpages as similar as possible to the targeted victim's.

\subsection{ML-based Anti-Phishing}
Generally, ML-based anti-phishing methods extract
a set of features of text, image, or URL information from DOM trees of
authentic and/or phishing websites.
A set of classification rules obtained from the learning algorithms are used for phishing
detection~\cite{VMA16}.
According to the existing ML-based anti-phishing solutions such as~\cite{zhang2007cantina,XHRC11,whittaker2010large,MAGSSA17,corona2017deltaphish,ubing2019phishing,ThomasGMPS11,ludl2007effectiveness},
we define a generic \emph{ML-based phishing classifier} $c$ as a tuple
\[c\triangleq (R,  \pi, \tau),\]
where $R$ is a finite set of classification rules,
$\pi$ is a function used to compute the final decision score of a given website, and
$\tau$ is the threshold.
Classification rules are described as follows:
\[\begin{array}{rl}
\mbox{Rule:}&  r={ (id_r,\ F_r, \ w_r)} \\
\mbox{Feature:}& f={ (n_f, \ v_f)} \\
\mbox{Node:}&  n ={\tt (n_{type}, \ n_{tag}, \ n_{children}, \ n_{v})	}
\end{array}\]
A classification rule $r$ consists of an identity $id_r$, a set of features $F_r$,
and a weight $w_r$ contributing to the final decision score.
We denote by $W=(w_r)_{r\in R}$ the set of weights.
A feature $f$ has a value $v_f$ which is computed from a set $n_f$ of DOM nodes of webpages.
The values of features in $F_r$ determine the value of $r$.
For example, a feature \texttt{PageHasForms} examines whether a webpage has a form node,
and its value is either $1$ or $0$.
A DOM node has a type ${\tt n_{type}}$ (e.g., element, attribute, text) and a tag name ${\tt n_{tag}}$ (e.g., ``div'').
An element node has children ${\tt n_{children}}$ that are attribute and text nodes.
An attribute or text node has a value ${\tt n_{value}}$.
For example, a ``type'' attribute node has the value ``password'' and a text node has the value ``Hello World!''.
In practise, classification rules are usually designed according to domain knowledge of experts, while their weights are trained from some
dataset using machine learning algorithms.

A classification rule $r$ is hit by a website $p$, denoted by $p\models r$,  only if all the values of its features in $p$ are non-zero.	
To check a website $p$, it computes the decision score ${\tt score}_c(p)$ based on weights of classification rules hit by $p$
using the function $\pi$.
It is considered to be a phishing one if the decision score exceeds a predefined threshold, i.e., ${\tt score}_c(p)\geq \tau,$
otherwise it is regarded as a legitimate one.
Although, classifiers may differ in concrete anti-phishing solutions,
our definition is generic and is able to capture cores of many published ML-based solutions, e.g.,~\cite{zhang2007cantina,XHRC11,whittaker2010large,MAGSSA17,corona2017deltaphish,ubing2019phishing,ThomasGMPS11,ludl2007effectiveness}.


\section{Threat Model and Overview}\label{sec:threat_and_overview}
In this section, we propose our threat model, and then introduce the overview of our study.

\subsection{Threat Model}
We assume that an attacker wants to craft a sample from a desired
phishing website that has the same functionalities and appearance, but is misclassified as benign by the target classifier;
and (s)he can query to the classifier as many times as possible.
This is feasible in practice, as limiting the number of query times
will restrict the usage of the classifier.

We consider three attack scenarios: white-, grey-, and black-box.
White-box means that the adversary has almost \emph{full} knowledge of the target classifiers (i.e., both classification rules and their weights).
Grey-box means that the adversary has access to classification rules, but not their weights, of the classifier under attack,
since the weights of classifiers may be frequently updated in practice.
For instance, Google retrains {\tt GPPF} daily on the server-side using samples from classification
data collected over the last three months and updates the weights of {\tt GPPF} on the client-side~\cite{whittaker2010large}.
By launching grey-box attacks, the adversary does not need to infer the weights of the classification rules, hence boosting attack efficiency.
In the more realistic black-box setting, the adversary does \emph{not} have any knowledge of the target classifiers.
In all three scenarios, the adversary does not know any information of the training dataset and training algorithm.

White- and grey-box attacks are feasible for classifiers that are deployed on the client-side~\cite{whittaker2010large}.
Indeed, thought labor intensive, classification rules (and their weights) can be explored through previously published work and many state-of-the-art techniques (e.g., reverse engineering~\cite{chikofsky1990reverse} and collision attacks~\cite{wang2005break}).
Consequently, the attacker may have full or partial knowledge of the classifier under attack, allowing (s)he to launch a white- or grey-box evasion attack on the target classifier.

We also assume that the classifier outputs the decision score of each input sample, which is
a widely used assumption even in black-box scenario~\cite{SL14,xu2016automatically,PMGJCS17,BHLS18}.
We will demonstrate that our approach is effective in transferability attacks,
allowing us to attack classifiers that only output category of each input instead of the decision score.

\subsection{Approach Overview}

  \begin{figure}[t]
  \centering
  \includegraphics[width=0.45\textwidth]{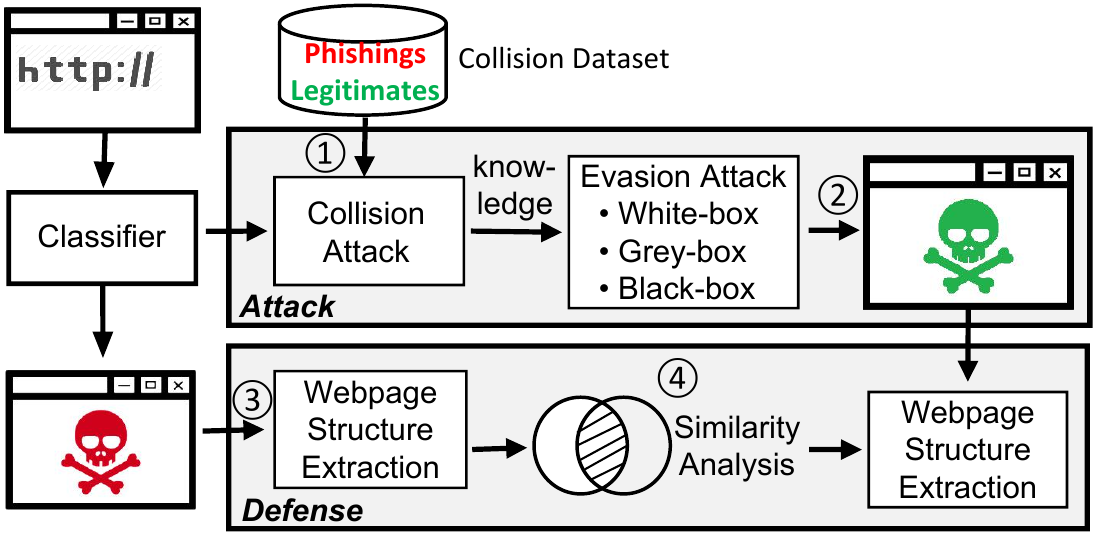}\vspace{-2mm}
   \captionof{figure}{Overview of our study.}
   \label{fig:overview}\vspace{-2mm}
 \end{figure}
Fig.~\ref{fig:overview} shows the overview of our work on attacks and defenses of
phishing classifiers: including advanced evasion attacks in white-, grey-, and black-box scenarios and an effective similarity-based method for evasion attack detection.
We first propose a sample-based collision to get the knowledge of classifiers such as classification rules and their corresponding weights (marked \ding{172}).
Second, given a desired phishing website that can be successfully classified as phishing by the classifier under attack, according to the knowledge the attacker obtained,
(s)he leverages evasion attacks to craft adversarial samples that mislead the classifier (marked \ding{173}).

In the defense part, when the original one is identified as phishing by the classifier, its webpage structures are extracted (marked \ding{174})
for evasion attack detection.
We propose an effective similarity-based method named \tool (marked \ding{175}),
which compares the similarity of DOM trees between an unknown one with recently detected phishing websites.
If the similarity exceeds a pre-defined threshold, we regard the input website
as a phishing one constructed by evasion attacks, but without invoking the ML-based classifier.

We further come up with two classification rule selection strategies for enhancing the robustness of the classifiers and increasing the attack cost.

\section{Advanced Evasion Attacks}\label{sec:attacks}

In this section, we propose a class of advanced evasion attacks under three different attack scenarios.
Given a desired phishing website $p$,
an  \emph{evasion attack}  on a target classifier $c=(R, \pi, \tau)$
is to craft an adversarial sample mutated from $p$ that can mislead the classifier $c$,
according to the knowledge of $c$ the attacker obtained.

To explain our approach, we will use \texttt{GPPF} as the target classifier,
which is trained by a gradient descent logistic regression
learning algorithm.
Recall that all the other tools we examined either are not publicly accessible or
cannot output decision score (cf. Section~\ref{Sec:Intr}), therefore, we demonstrate our generic methods on the state-of-the-art classifier \texttt{GPPF}.
In \texttt{GPPF} $c_{{\tt GPPF}}=(R, \pi, \tau)$,
$R$ consists of 2,130 classification rules that rely upon 995 features in total;
the decision score function $\pi$ is defined as: \[\pi(x)=\frac{e^x}{1+e^x},\] where
$x=w_{r_1}+\sum_{i=2}^{2130}(w_{r_i} \cdot \prod_{f\in F_{r_i}}v_f)$
and the threshold $\tau$ is $0.5$. 

\subsection{Sample-based Collision Attack}
\label{sec:collision-attack}
To lanuch a grey- or white-box attack on a target classifier,
it is necessary, but non-trivial, to obtain full or partial information of the classifier.

 \begin{table}[t]\footnotesize
 	\centering
	\caption{Dataset used in this work} \label{table:dataset}\vspace{-2mm}
\begin{tabular}{ccc}  \toprule
		{\bf Dataset Types} & {\bf Number} & {\bf Source} \\ \midrule 
		Phishing URLs & 2,566,834 & \textsc{PhishTank} \\
		Phishing Websites & 3,000 & \textsc{PhishTank} \\ 	\midrule  
		Legitimate URLs & 15,000 &  \textsc{PhishNet}\\
		Legitimate Websites & 15,000 &  \textsc{PhishNet}  \\ \bottomrule
	\end{tabular}\vspace{-3mm}
 \end{table}

\subsubsection{Data preparation}
As shown in Table~\ref{table:dataset}, we collect phishing websites from \textsc{PhishTank}~\cite{phishtank}, 
and legitimate websites from
\textsc{PhishNet}~\cite{phishnet}, as subjects to infer classification rules.
We only collect URLs of inactive phishing websites.
Finally, we collected 3,000 phishing websites and 2,566,834 URLs from \textsc{PhishTank} and 15,000 legitimate websites with their corresponding webpages from \textsc{PhishNet}.
%

Specific data should be extracted from the collected websites in order to launch collision attacks,
as features may differ in classifiers.
As aforementioned, we demonstrate our methods using \texttt{GPPF} as the target classifier.
\texttt{GPPF} consists of three categories of features: URL-related, Dom-related, and term-related features, as shown in Table~\ref{table:encrypt features}.
URL-related features come from URLs, containing 4 types of features.
Dom-related features come from links and action fields (e.g., form element) in webpages, containing 2 types of features.
Term-related features are collected from texts in webpages, including one type of features (i.e., \texttt{PageTerm}).
We extract all links and texts from the collected websites and URLs as the dataset for inferring classification rules of \texttt{GPPF}.



\subsubsection{Collision attacks}
We demonstrate how to infer classification rules in \texttt{GPPF} by leveraging the collected dataset.
The main goal is to identify the features used in classification rules.
\texttt{GPPF} uses the \texttt{SHA-256} hash algorithm to protect these features.
Therefore, we compute hash value of each feature in the dataset using the \texttt{SHA-256} hash algorithm.
We then compare its hash value with the hard-coded hash values in the source code of \texttt{GPPF}. If it matches one of the hard-coded hash values in \texttt{GPPF},
then the feature is identified. 
It remains to understand the semantics of the identified features,
which is required to launch white- and grey-box attacks.
However, it is difficult to recognize it from the feature names. For example, it is hard to guess the semantics of the feature {PageLinkDomain=$*$},
which actually means that ``a'' element with the ``href'' attribute whose value is string (i.e., $*$ denotes wildcard).
To address this problem, we manually extract the semantics of each feature from the code of \texttt{GPPF}.
The results are shown in Table~\ref{table:Deletion and addition for GPPF features}, which will be explained in detail in Section~\ref{explain_table3}.
{\textit{This approach is not limited to the \texttt{SHA-256} hash algorithm or \texttt{GPPF}, instead,
we believe it could be used for other hash algorithms and client-side classifiers.}}

  \begin{table}[t]\footnotesize
  	\centering
	\caption{Extracted features in \texttt{GPPF}} \label{table:encrypt features}\begin{tabular}{cc}
		\toprule
		{\bf Feature Categories} & {\bf Feature Types} \\ \midrule
		\multirow{2}{*}{\bf URL-related} & UrlTld=$*$, UrlDomain=$*$ \\
		& UrlOtherHostToken=$*$, UrlPathToken=$*$ \\ \midrule  
		\multirow{1}{*}{\bf Dom-related} &  PageActionURL=$*$, PageLinkDomain=$*$ \\ \midrule
		{\bf Term-related} & PageTerm=$*$ \\ \bottomrule
	\end{tabular}
 \end{table}

\subsection{Mutation Mechanisms}\label{sec:mutation-mechanism}
We propose three mutation mechanisms for white-, grey- and black-box evasion attacks,
which craft adversarial samples by leveraging mutation techniques.
During mutation, we always preserve the appearance and
functionalities of the websites. This is achieved by carefully designing mutation operations for implementing our mutation mechanisms.
Since changing URLs of websites is very easy for an attacker in real scenarios,
but it is difficult to automatically craft adversarial samples via mutating URLs in order to maintain similarity to victim's URLs.
Therefore, we do not try to modify URLs in our evasion attacks.


\subsubsection{\bf White-box Attacks}
To evade a classifier with its full knowledge,
we should mutate a sample starting from a phishing website by deleting or adding classification rules until
the decision score is less than the threshold.
The challenge is how to choose such classification rules.
Since classification rules can contribute to the decision score either positively or negatively,
we partition the classification rules into two sets:
\begin{itemize}
  \item $R^+=\{r\in R\mid w_r>0\}$
the set of \emph{positive} classification rules indicating when being added the decision score will increase,
  \item  $R^-=\{r\in R\mid w_r<0\}$
the set of \emph{negative} classification rules indicating when being added the decision score will decrease.
\end{itemize}


We address the challenge by deleting positive classification rules and add negative classification rules, so that the decision score finally becomes less than the threshold.
We leverage a greedy algorithm to choose which classification rules to add or delete aiming at maximally
decreasing the decision score at each mutation step.
It is worth noting that once a feature is deleted, all the classification rules relying on this feature are also deleted.
On the other hand, in order to add a  classification rule, all the features relied by this classification rule should be added.
Therefore, we regard each feature of negative classification rules as a mutation ``atom'' during deletion operation,
and regard each positive classification rule as a mutation ``atom'' during addition operation
instead of individual features.
At each mutation step, features and classification rules are selected based on
their influence on the decision score.

For each feature $f$ that appears in the website $p$,
we define the influence $\delta(p,f)$ of $f$ as 
\begin{center}$\delta(p,f)\triangleq\sum_{r\in R(f)\cap p\models r}(w_r\cdot\prod_{f'\in F_r}v_{f'}),$\end{center}
where $v_{f'}$ denotes the value of feature $f'$ in $p$ and $R(f)$ denotes the set of classification rules that 
rely upon $f$.
Intuitively, $\delta(p,f)$ is the diff of weights
after deleting the feature $f$.

To add one negative classification rule $r$, all the features of $r$ should be added,
by which more than one classification rules may be added, hence adding the features of one classification rule may result in the addition of more than one classification rules.
Therefore, we define the influence $\delta(p,r)$ of a classification rule $r$ as
\begin{center}$\delta(p,r)\triangleq \sum_{r'\in R\wedge p\not\models r'\wedge F_{r'}\subseteq F_r} (w_{r'}\cdot \prod_{f'\in F_{r'}}v_{f'}),$\end{center}
indicating the diff of weights
after adding all the features of the classification rule $r$.

Now, we describe the procedure of our white-box attack.
For a given phishing website $p$, we iteratively mutate $p$ as follows until its decision score
becomes less than the threshold: 

\begin{enumerate}[noitemsep,topsep=0pt,leftmargin=*]
	\item if there exists a feature $f$ of a positive classification rule such that $\delta(p,f)>0$, $\delta(p,f)\geq \delta(p,f')$ for all other features $f'$ of positive classification rules,
	and $\delta(p,f)\geq |\delta(p,r)|$ for all negative classification rules $r$, then delete $f$ from the website $p$; otherwise goto step 2.
	\item if there is a negative classification rule $r$ in the website $p$ such that $\delta(p,r)<0$ and $\delta(p,r)\leq \delta(p,r')$ for all other negative classification rules $r'$, then add
	the classification rule $r$ (i.e., all the features of $r$) into $p$; otherwise terminate.
\end{enumerate}

\subsubsection{\bf  Grey-box Attacks}
In grey-box scenario, the attacker knows classification rules, 
but not weights of the classification rules.
Therefore, it is impossible to choose the best
features or classification rules so that the decision score can be reduced maximally at
each mutation step.
For efficiency consideration,
based on our insights of ML-based classifiers (cf. Section~\ref{sec:expcollision-attack}), we first iteratively delete features that can reduce the decision  score;
if the decision score is still no less than the threshold after deleting all such features,
we add other classification rules of the target classifier that do not exist in the website and check which classification rule can reduce the decision score 
until no more classification rules can be added.
We stop deleting/adding once the decision  score becomes less than the threshold.
Since we do not know the weights of rules, feature deletion and addition operations may be interleaved during mutation.


\subsubsection{\bf Black-box Attacks}
In black-box scenario, the attacker neither knows classification rules nor their weights.
Hence, it is impossible to directly delete/add classification rules or features during mutation.
To overcome this problem, we iteratively add and/or modify DOM nodes, instead of classification rules or features.
According to our insight of existing classification rules in ML-based classifiers,
we find that:
(1) deleting one feature leads to the deletion of all rules that rely on this feature,
and (2) to add a rule, all the features of this rule should be added.
Therefore, we first iteratively modify DOM nodes.
If the decision score becomes less than the threshold, we stop mutating.
Otherwise, we iteratively and randomly add nodes that are extracted from legitimate websites, but do not exist in the phishing website, to reduce the decision  score.
If the decision  score becomes smaller, we continue mutating the current website.
Otherwise, we continue mutating the previous one.
We will elaborate details in the next subsection.
All these operations are carefully designed in order to preserve
functionality and appearance of the phishing webpage, but without using the obtained features
and weights. Details

\subsection{Mutation Operations}\label{mutation-operations}
According to the above mechanisms,
our white- and grey-box attacks need to delete and/or add features.
These operations are achieved by manipulating DOM nodes,  the same as in black-box scenario.
We extract the DOM nodes related to a feature, and
add/delete this feature by manipulating these nodes.
For example, consider the feature named \texttt{PageHasPswdInputs} in \texttt{GPPF},
here we assume that ``\texttt{PageHasPswdInputs} $=$ True'' in the sample website.
To delete this feature, we identify the nodes (i.e., \texttt{input}) related to this feature, then modify these nodes
to break ``\texttt{PageHasPswdInputs} $=$ True''.
Hence, manipulating DOM nodes is a key operation to achieve our evasion attacks.
However, directly manipulating DOM nodes will
destruct appearance and/or functionalities of webpages, which are determined by the DOM tree and their nodes. 
To address this challenge, we introduce two mutation operations: node ``addition'' and node ``modification'',
which respectively adds and modifies DOM nodes so that
(G1) the values of features in positive classification rules become zero,
or (G2)  the values of features in negative classification rules become positive,
but without destructing the appearance and functionalities.

\subsubsection{\bf Node Modification}	
{To achieve G1 or G2,} we illustrate node modification for different types of nodes (i.e., attribute nodes and text nodes) as follows.


\smallskip
\noindent
{\bf Attribute nodes}.
Attribute nodes are modified by replacing them with other grammars, so that these nodes will not be the features of the classifier.
For instance, we can use ``event'' in \texttt{HTML} to achieve this goal.
 \texttt{HTML4} and \texttt{HTML5} support the event trigger actions in a browser,
and the effect is the same as executing a script when a user clicks on an element. In addition, in order to ensure the same appearance, we also need to add the style attribute.
For instance, the attribute ``type'' whose value is ``submit'' in the button element can be modified, as shown below.
\begin{lstlisting}[backgroundcolor=\color{gray!20}]
<style type="text/css">button[type=submit]{height:38px}</style>
<button type="submit"></button> //original\end{lstlisting}
%
%
We replace the attribute ``type'' with value ``submit'' by leveraging the ``onclick'' event,
parse the ``css'' file and ``style'' tags to find the styles defined for the button,
and add these styles to the button so that the appearance of the button remain unchanged. The crafted button is shown below.
\begin{lstlisting}[backgroundcolor=\color{blue!20}]
<button style='height:38px;'onclick="this.type='submit'";></button> //crafted
\end{lstlisting}
It is very difficult for the users to perceive such changes since the two actions do not destruct the appearance and functionality of webpages.
Table~\ref{table:deletable_attributes} lists the function-related attribute nodes which can be modified.
We remark that
not all attribute nodes can be modified, as some of them are related to webpage appearance (e.g., ``align'' attribute in a ``p'' element and ``background'' attribute in a ``body'' element)
unless the appearance can be imitated by changing ``style'' attribute. 
For example, we cannot modify the attribute ``type'' whose value is ``radio'' in ``input'' element, because its appearance cannot be imitated by changing ``style'' attribute.

\begin{table}\footnotesize
	\centering\setlength{\tabcolsep}{1pt}
	\caption{Modifiable function-related attributes where $*$ refers to the ``type'' attribute in ``input'' element, which can be deleted only when its value is ``reset'', ``button'', or ``password''.}
	\label{table:deletable_attributes}\vspace{-2mm}
	\scalebox{0.95}{
	\begin{tabular}{ccc}
		\toprule
		{\bf Element} & {\bf Function-related Attributes} & {\bf Event} \\ \hline  
		button & \begin{tabular}[c]{@{}c@{}}form, formaction, formenctype, \\ formmethod, formnovalidate,  formtarget,\\  name, type, value\end{tabular} & onclick \\ \midrule
		input & \begin{tabular}[c]{@{}c@{}}accept, submit, text, type$*$, step, size, \\  required, name,  placeholder, min, max,\\  formtarget, formnovalidate, formmethod, \\ formenctype,  formaction, form, accept\end{tabular} & onfocus \\ \midrule  
		form & \begin{tabular}[c]{@{}c@{}}action, enctype, method,  name, novalidate, target\end{tabular} & oninput \\ \midrule
		a & \begin{tabular}[c]{@{}c@{}}download, href, hreflang,  media, rel, target, type\end{tabular} & onclick \\ \midrule 
		table & summary & onmousemove \\
		\bottomrule
	\end{tabular}}\vspace{-3mm}
\end{table}



\smallskip
\noindent
{\bf Text nodes}.
{Classifiers usually check text through string comparison.}
Thus, we can modify text nodes by inserting an invisible (e.g., the  zero-width space ``\&\#8203;'') character, which looks like that it does not exist.
For example, a text node ``Hello World!'' in a ``p'' element, as shown below
\begin{lstlisting}[backgroundcolor=\color{gray!20}]
<p>Hello World!</p> //original\end{lstlisting}
It can be modified by inserting the character ``\&\#8203;'' between
``l'' and ``o'',  displayed the same as the original one, resulting in the following code:
\begin{lstlisting}[backgroundcolor=\color{blue!20}]
<p>Hell&#8203;o World!</p> // crafted
\end{lstlisting}



\subsubsection{\bf Node Addition}
In order to preserve the appearance and functionalities of webpages,
we demonstrate node addition by adding
\emph{invisible} element nodes, i.e., with no size and color.
We remark that the parent of an attribute or text node is an element node.
If the element node is invisible, then all its children are also invisible.
Therefore, to add attribute and text nodes,
we add them as children of an invisible element node.

However, in black-box scenario, we do not know any information of the classifier and thus cannot determine which node to add in order to hit negative classification rules after node modification.
We assume that there are more negative classification rules in legitimate websites than in phishing ones.
Therefore, based on this assumption,
we collect three types of nodes (i.e., element nodes, attribute nodes, and text nodes) from the collected legitimate websites.
After that, we randomly select nodes from the collected data
and add them into the target website.
To avoid adding useless nodes in black-box scenario, we come up with a \textit{rollback mechanism}.
Specifically, if the decision score does not decrease after
adding a specified number (e.g., $3$) of nodes,
we rollback to the previous sample and continue to add other nodes.
Otherwise, we continue adding nodes into the latest one.

\vspace{2mm}
\noindent\fbox{
\parbox{0.95\linewidth}{
In summary, all nodes can be added into the webpages, but only attribute and text nodes can be modified according to mutation operations.
}}
\vspace{2mm}

Though our node modification and addition are simple, they are effective in our experiments.
It might be able to improve resilience of classifiers by proposing new classification rules following disclosure of our findings,
we believe more sophisticated methods (e.g., script and Shadow DOM~\cite{HAJSS14}) could be leveraged to achieve node modification and addition,
which are widely used in legitimate websites,
but very difficult to analyze statically.

\subsection{Concretizing Our Attacks on GPPF} \label{explain_table3}
In this section, we show how to concretize our attacks on the classifier \texttt{GPPF}.
Table~\ref{table:Deletion and addition for GPPF features} lists all features excluding URL-related features in \texttt{GPPF}.
As aforementioned, all the features can be added, while only some features can be deleted.
For example, the feature \texttt{PageHasForms} denotes whether a webpage has a ``form'' node.
The ``form'' node is an element node which cannot be modified.
We cannot delete features 4 and 5 neither, as they are appearance-related attribute nodes.
Features 6, 7, 8 and 9 representing the frequency of special links can be deleted by adding links whose domain is internal.
When the frequency value is lower than a predefined threshold in \texttt{GPPF}, \texttt{GPPF} cannot detect such features.
We cannot delete script-related features 10 and 11, as deleting them may affect the functionalities of websites.

Features 2, 3, 12 and 13 can be deleted by using the attribute nodes (e.g., \texttt{event}) which implement the same functionality,
but making them undetectable by the classifier. For example, feature 3 indicates whether the website has an ``input'' element whose ``type'' is ``password'', as shown below:
\begin{lstlisting}[backgroundcolor=\color{gray!20}]
<style type="text/css">input[type=password]{width:8px;}</style>
<input type="password" /> // original\end{lstlisting}

After modifying it using the  \texttt{event} attribute node,
it becomes
 \begin{lstlisting}[backgroundcolor=\color{blue!20}]
<input style="width:8px" onfocus="this.type=`password'" /> //crafted
\end{lstlisting}
Thus, the classifier cannot detect this feature. For feature 14, we can delete it directly since it is a term-related feature.

\section{Defense Methods}\label{sec:defesne}
In this section, 
%
we first propose a similarity-based defense method, named \tool, to detect evasion attacks. 
\tool addresses the scenario where a phishing website is already detected by a classifier,
and the attacker is trying to mutate it to mislead the classifier. 
%
We then propose two classification rule selection strategies to increase
attack cost, hence enhancing the robustness of the classifier.


\subsection{Similarity-based Evading Detection}
Our detection is based on the following two observations of mutation based evasion attacks:
(1) adding or deleting features do change the structures of the webpages,
but, (2) the change is limited.
Therefore, if a website has a high ``similarity'' compared with a
recently detected phishing one, we regard it as an adversarial sample crafted by evasion attacks.
We remark that our approach aims at detecting evasion attacks rather than general
phishing attacks. It can be deployed with a ML-based classifier to prevent from mutation based evasion attacks.

\subsubsection{\bf DOM Tree Similarity (Baseline)}
We first propose to compute the similarity of DOM trees as baseline. 
We denote by $T$ and $T'$ two DOM trees of a recently detected phishing website and an unknown website.
If the similarity between $T$ and $T'$ is high, we conclude that ${T}'$ is mutated from $T$, thus
successfully identify the crafted phishing website
without using the machine learning-based classifiers.

In detail, we traverse the DOM trees $T$ and $T'$ using a breadth first search algorithm
and compute the hash values of the element nodes in each layer.
As there are two types of nodes (i.e., text nodes and attribute nodes) that rely on an element node,
thus, to measure the similarity of two element nodes, we compare both the hash values of their text and attribute nodes.
{We only consider the similarity of two elements with the same tag name. If two elements have distinct tag names, they are regarded as different elements.  The similarity of elements in the $i^{th}$ layer is defined as:}
\begin{center}
	$ \tt similarity_{ele}(E_i,E_i') \triangleq\frac{|a_1^i\cap a_2^i|}{|a^i_1\cup a^i_2|}+\frac{|t^i_1\cap t^i_2|}{|t^i_1\cup t^i_2|}$
\end{center}
%
where $a^i_1$ (resp. $a^i_2$) denotes the set of attribute nodes belonging to the element node $E_i$ (resp. $E_i'$),
$t^i_1$ (resp. $t^i_2$) denotes the set of text nodes belonging to the element node $E_i$ (resp. $E_i'$).
Finally, we compute the proportion of the common nodes ${\tt comm}_i$ with the same hash values against
all the nodes of the current layer $i$ in two DOM trees.
The similarity ${\tt similarity_{tree}(T,T')}$ between $T$ and $T'$ is defined as the average common nodes of each layer, i.e.,
\begin{center}
${\tt similarity_{tree}(T,T')}\triangleq \sum_{i=1}^{m}\frac{\ell_i}{m}$
\end{center}
where $m$ is the maximal number of layers in the DOM trees,
\begin{center}
$\ell_i\triangleq\frac{{\tt comm_i}}{|n^i_1\cup n^i_2|}$ {and}
	  ${\tt comm_i} \triangleq \sum_{j=1}^{{|n^i_1\cup n^i_2|}} \tt similarity_{ele}(E_j,E_j')$,
\end{center}
%
  $n^i_1$ (resp. $n^i_2$)
denotes the set of element nodes at the $i^{th}$ layer of $T$ (resp. $T'$).

\begin{table*}\footnotesize
	\centering\setlength{\tabcolsep}{3pt}
	\caption{Dom-related and Term-related features in \texttt{GPPF} with their semantics and attributes, where ``*'' refers to wildcard}
	\label{table:Deletion and addition for GPPF features}

	\begin{tabular}{lcccp{340pt}} \toprule
		 & {\bf Feature Types} & {\bf Del} & {\bf Add} & {\bf Semantics} \\ \midrule 
		1 & PageHasForms & \xmark & \cmark & Whether there is a form element in the webpage. \\ \midrule 	
		2 & PageHasTextInputs & \cmark & \cmark & Whether there is an input element whose type is ``text''. \\ \midrule 
		3 & PageHasPswdInputs & \cmark & \cmark & Whether there is an input element whose type is ``password''. \\ \midrule
		4 & PageHasRadioInputs & \xmark & \cmark & Whether there is an input element whose type is ``radio''. \\ \midrule 
		5 & PageHasCheckInputs & \xmark & \cmark & Whether there is an input element whose type is ``checkbox''.\\ \midrule
		6 & PageExternalLinksFreq & \cmark & \cmark  &Computed as $\frac{\tt \#external\_links}{\tt \#total\_links}$, where ${\tt \#total\_links}$ is the No. of ``href'' attributes in ``a'' elements and ${\tt external\_links}$ is the No. of ``href'' attributes in ``a'' elements whose domains are external.
		\\ \midrule
		7 & PageActionOtherDomainFreq & \cmark & \cmark & Computed as $\frac{\tt \#actions_{other\_domain}}{\tt \#total\_actions}$,
		where ${\tt \#total\_actions}$ is the number of the ``action'' attributes in the form element and
		${\tt \#actions_{other\_domain}}$ is the number of ``action'' attributes whose domains are external.  \\ \midrule
		8 & PageSecureLinksFreq & \cmark & \cmark &Computed as $\frac{\tt \#secure\_links}{\tt \#total\_links}$, where ${\tt \#total\_links}$ is the number of ``href'' attributes in ``a'' elements and ${\tt \#secure\_links}$ is the number of ``href'' attributes whose scheme is https in ``a'' elements. \\ \midrule  
		9 & PageImgOtherDomainFreq & \cmark & \cmark &Computed as $\frac{\tt \#imgs_{other\_domain}}{\tt \#total\_imgs}$, where ${\tt \#total\_imgs}$ is the number of ``img'' elements in the webpage. ${\tt \#imgs_{other\_domain}}$ is the number of ``srr'' attributes whose domains are external. \\ \midrule
		10 & PageNumScriptTags\textgreater{}1 & \xmark & \cmark & Whether the number of ``script'' tag is greater than $1$.\\ \midrule  
		11 & PageNumScriptTags\textgreater{}6 & \xmark & \cmark & Whether the number of ``script'' tag is greater than $6$. \\ \midrule
		12 & PageActionURL=$*$ & \cmark & \cmark 	&  Whether there is a form element with the ``action'' attribute whose value is a string. \\ \midrule  
		13 & PageLinkDomain=$*$ & \cmark & \cmark & Whether there is an ``a'' element with the external ``href'' attribute whose value is a string. \\ \midrule
		14 & PageTerm=$*$ & \cmark & \cmark &Whether there is a text node containing a string.  \\ \bottomrule
	\end{tabular}
\end{table*}

\subsubsection{\bf Personalized Similarity (\tool)}
The DOM tree similarity method (baseline) is effective to detect most crafted samples,
but failed on several cases, e.g.,  adversarial samples crafted by the grey-box attack.
After an in-depth analysis, we found there are four types of features related to ``frequency'',
{namely, features at No. 6-9 in Table~\ref{table:Deletion and addition for GPPF features}.}
In order to evade classifiers, it requires to add or modify a large number of nodes,
which may significantly reduce the DOM tree similarity, occurring in our grey-box scenario.
In addition, if the attacker obtains the knowledge of our DOM tree similarity method,
(s)he may add more irrelevant nodes or attributes, invisible elements, and layers to reduce the similarity. However, it is difficult to directly remove the nodes {and attributes} without destructing the appearance and functionalities.
Therefore, we propose a \emph{personalized similarity method}, named \tool, that is defined as follows:
%
\begin{center}\small
	$  \ \tt similarity_{ele\_\tool}(E_i,E_i') \triangleq\frac{|a_1^i\cap a^i_2|}{|a^i_1|}+\frac{|t^i_1\cap t^i_2|}{|t^i_1|}$,\\
	${\tt similarity_{\tool}(T,T') \triangleq \sum_{i=1}^{m}\frac{\ell_i}{m}}, \ \ell_i\triangleq\frac{{\tt comm_i}}{|n^i_1|}$ and\\
 ${\tt comm_i}\triangleq \sum_{j=1}^{{|n^i_1|}} \tt similarity_{ele\_\tool}(E_j,E_j')$
\end{center}
where $\tt similarity_{ele\_\tool}(E_i,E_i')$ is the similarity of elements in the $i^{th}$ layer by using \tool, $m$ is the number of layers in the DOM tree $T$,
and $n^i$ denotes the set of element nodes at the $i^{th}$ layer in $T$.
In particular, to defend against attacks that insert layers into the DOM tree during mutation, \tool continues comparing the next layer of ${T}'$ until the two layers are similar. If all the remaining layers are not similar enough, we just rollback to compute the similarity of elements in the same layer.

\begin{figure}[t]
  \centering
  \includegraphics[width=0.5\textwidth]{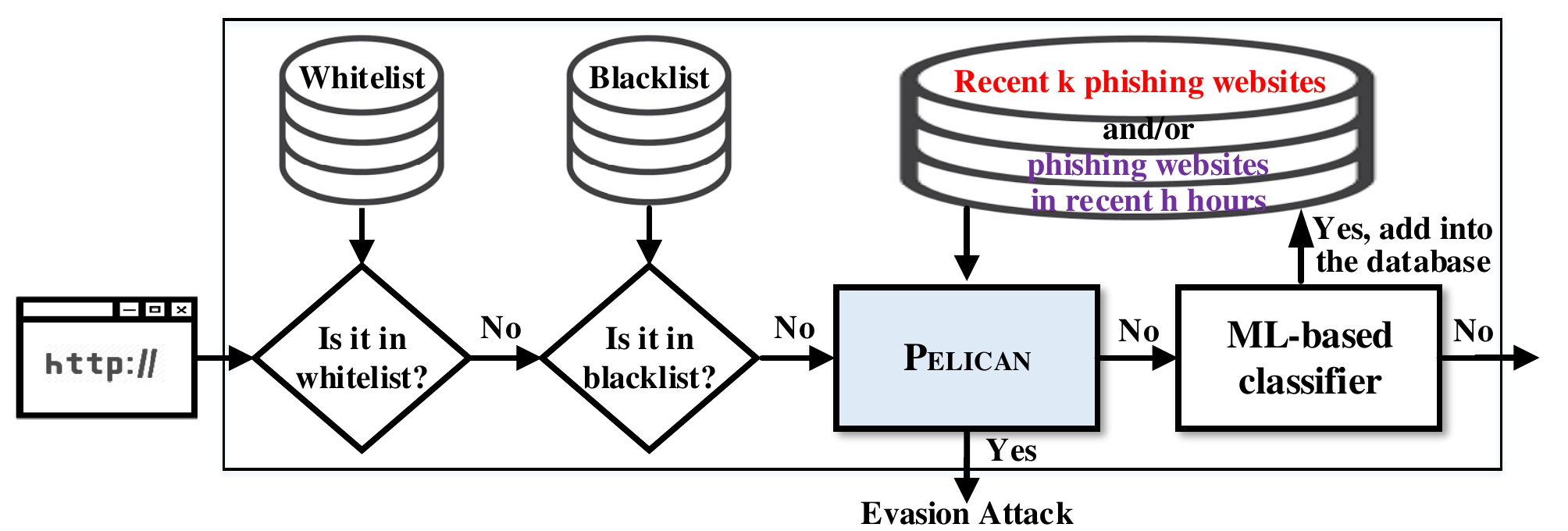}
  \caption{Application scenario in practice.}\label{fig:defense-app}
\end{figure}

\subsubsection{\bf Application Scenario in Practice}
Fig.~\ref{fig:defense-app} depicts the
application scenario of our defense method \tool in practice.
For each website, the system first queries
whitelist and blacklist to quickly identify
legitimate and known phishing website in a standard way.
{Whitelist and blacklist can be customized by users or use third-party resources~\cite{safebrowsingAPI,corona2017deltaphish,phishtank,phishnet}.}
If the website is neither in whitelist nor blacklist,
then \tool is applied to check whether
the website is crafted by an evasion attack.
If it is an evasion attack, a phishing warn alerts
without passing to the ML-based classifier.
Otherwise,  ML-based classifier is used to
check the website. The website is added into the phishing website database
for \tool if it is classified as a phishing one.
To avoid storing all the detected phishing websites,
we only need to store either recent $k$ phishing websites
and/or the phishing websites detected in recent $h$ hours.
By doing so, \tool can prevent from querying directly and efficiently to the ML-based classifiers by the attackers,
but still detect phishing attacks.


\subsection{Classification Rule Selection}
In order to enhance the robustness of the classifiers, we introduce two classification rule selection strategies to increase attack cost.

\smallskip
\noindent{\bf Negative subset classification rule pruning}.
Given two classification rules $r$ and $r'$,
we say $r'$ is a \emph{sub-rule} of $r$,
if the set of features of $r$ subsumes the one of $r'$.
Such subset classification rules decrease the attack overhead, because if the attacker has inferred the classification rule $r'$,
then the classification rule $r$ as well.
%
%
Attackers can improve the attack effectiveness and efficiency by leveraging negative
subset classification rules. 
When the attacker adds all the features of a negative classification rule $r$ into a website, both classification rules $r$ and $r'$ will be added into the sample.
Therefore, the mutation can maximally reduce the decision score by adding all features of $r$.
To enhance the robustness of the classifiers, it is better to prune subset classification rules, which is demonstrated by our experiments (cf. Section~\ref{sec:expsubsetruleimprove}).

\smallskip
\noindent
{\bf Single classification rule pruning}.
A classification rule is \emph{single} if it relies on features that are not relied by any other classification rules.
Single classification rules are completely independent to other classification rules,
deleting or adding them will not affect any other classification rules.
Therefore, attackers can aggressively add all negative single classification rules if they are addable, which can definitely reduce the decision score.
In addition, in white-box scenario, attackers can also directly delete all positive single classification rules.
Addition and deletion of single classification rules are very efficient and do not have any side effect to other classification rules.
Therefore, single classification rules have a potentially severe hazard to classifiers, and should be pruned in order to enhance the robustness
of the classifiers.
This also has been confirmed by our experiments (cf. Section~\ref{sec:expsingleruleimprove}).

	\section{Experiments}\label{sec:experiments}
In this section, we first introduce the dataset and experimental settings,
then demonstrate our sample-based collision attack for inferring classification rules,
and finally, evaluate our proposed evasion attacks
and defense methods for evading detection and robustness improvement of the classifier.

	\subsection{Dataset and Experimental Settings}
	
\noindent{\bf Dataset.}
We use two datasets (cf. Table~\ref{table:dataset}),
for evaluating our collision attack.
We tested 3,000 phishing websites using the classifier {\tt GPPF}. Among them, 305 can be detected by  {\tt GPPF} (i.e., $score\geq 0.5$).
We use these 305 phishing websites as the subjects to evaluate evasion attacks and defense methods.


\smallskip
\noindent{\bf Settings.}
As mentioned in Section~\ref{Sec:Intr},
we examined several academic and industrial ML-based tools, all the other tools (except \texttt{GPPF}) either are not publicly accessible or cannot output decision score for each input which is required even in our black-box scenario. Thus, we demonstrate our attacks
on the state-of-the-art ML-based classifier \texttt{GPPF}.

We use \textsc{depot\_tools}~\cite{depottools19} to obtain the code of \texttt{Chromium} (version: 72.0.3626),
and use \texttt{ninja}~\cite{ninja} to compile \texttt{Chromium}.
We target the classifier \texttt{GPPF} which is deployed in \texttt{Chromium}.
To obtain the decision  score of each website, we added an instruction which prints the return value of the function {\tt ComputeScore} in the file {\tt scorer.cc}.
We run \texttt{Chromium} in UI mode for our experiments.
Our experiments are carried out on a server (i.e., Ubuntu 16.40 or Windows 10) with 32 GB Memory and 1 TB hard disk.
	
To demonstrate that our black-box attack is generic and effective for other practical ML-based classifiers,
we launch transferability attacks on the industrial tool {TrafficLight}, coming from BitDefender,
a popular security company. It is able to detect phishing webpages and part of the processing is
done in the cloud with some intelligent engines.
Recall that BitDefender TrafficLight is proprietary without any publicly
available information about the internal design and implementations,
hence completely black-box.
However, only 32 out of 305 seeds
can be detected by BitDefender {TrafficLight}.
Thus, we use all the crafted adversarial samples ($32\times 3$) from these 32 phishing websites as subjects.
We also tried to launch transferability attacks on other anti-phishing tools, but failed to get their classification results \emph{automatically}.

\begin{table}[t]\footnotesize
	\centering\setlength{\tabcolsep}{3pt}
	\caption{Comparison of inferred features}
	\label{table:feature decryption}
	\scalebox{0.9}{
	\begin{tabular}{@{}cccc|cc@{}}
		\toprule
		\begin{tabular}[c]{@{}c@{}}{\bf Feature} \\ {\bf Categories}\end{tabular} &
		\begin{tabular}[c]{@{}c@{}}{\bf \#Total} \\ {\bf Features}\end{tabular} & \begin{tabular}[c]{@{}c@{}}{\bf \#Inferred} \\ {\bf Features}\end{tabular} & \begin{tabular}[c]{@{}c@{}}{\bf \#Inferred} \\ {\bf in~\cite{liang2016cracking}}\end{tabular} & {\bf Time (h)} & {\bf Time in~\cite{liang2016cracking} (h)} \\ \midrule
		{\bf URL-related} & \multirow{2}{*}{563} & \multirow{2}{*}{491} & \multirow{2}{*}{426} & \multirow{2}{*}{0.68} & \multirow{2}{*}{0.58} \\
		{\bf Dom-related} &  &  &  &  \\ \midrule
		{\bf Term-related} & 432 & 387 & 375 & 11.66 & 25.13 \\ \midrule
		{\bf Total} & 995 & \multicolumn{1}{>{\columncolor{mygray}}c}{878} & {801} & \multicolumn{1}{>{\columncolor{mygray}}c}{12.34} & 25.71 \\ \bottomrule
	\end{tabular}}
\vspace{-3mm}
\end{table}
	
	\subsection{\bf Evaluation of Collision Attack}
\label{sec:expcollision-attack}
To evaluate the effectiveness of our \textit{sample-based collision attack},
we demonstrate on the classifier \texttt{GPPF}. 
We leverage our collected phishing and legitimate dataset from \textsc{PhishTank} and \textsc{PhishNet} in Table~\ref{table:dataset}.
We identify 995 hashed features after analyzing \texttt{Chromium}, as shown in Table~\ref{table:feature decryption}.
Our collision attack is able to successfully infer 878 features (88.24\%).
We compare our sample-based collision attack with the one proposed by Liang et al.~\cite{liang2016cracking}.
{Liang et al. use URLs to infer URL- and Dom-related features,
 alphabets and full-text corpora with seven languages to infer term-related features.
As their tool and sources of datasets are not available for ethical consideration,}
we list the results from Liang et al.~\cite{liang2016cracking}.
Our collision attack inferred more features (878 vs. 801) using less time (12.34 h vs. 25.71 h)
than the one in~\cite{liang2016cracking}.

In detail, our collision attack successfully inferred
65 more (491 vs. 426) URL-related and Dom-related features in 0.68 hour,
and 12 more (387 vs. 375) term-related features in less time
(11.66 h vs. 25.13 h), owing to our novel collision attack.
We also identified a new type of Dom-related features called  ``{\tt PageActionURL}'' which was not mentioned in \cite{liang2016cracking}.
We infer more term-related features than \cite{liang2016cracking},
e.g.,
term-related features
(1) containing numbers,
(2) involving Arabic,
or (3) using degree as value such as ``$1^\circ$''. 
{\bf For ethical consideration, we do not provide details of concrete features.}
Those features cannot be easily inferred using corpora, as done in~\cite{liang2016cracking}.
Our collision attack successfully inferred them because those terms are easy to be found in real-world websites.
In terms of time performance, to infer the term-related features,
\cite{liang2016cracking} uses words with lots of noise, thus needs much more time than ours.
In summary, our collision attack can crack more features using less time than that in~\cite{liang2016cracking}.



\begin{table}[t]\footnotesize
\centering
\caption{Statistic of inferred classification rules}\label{table:rule decryption}
\scalebox{0.9}{
\begin{tabular}{c|c|c|c|c|c}
\toprule
{\bf Rule Weight} & {\bf \#Total}  & {\bf \#Addable}   & {\bf Rule Weight} &{\bf \#Total} & {\bf \#Deletable} \\ \midrule \rowcolor{gray!20}
				{[}-6.0,-7.0) & 2 & 1 & {[}6.0,7.0) & 6 & 4  \\
				{[}-5.0,-6.0) & 2 & 0 & {[}5.0,6.0) & 13 & 8  \\  \rowcolor{gray!20}
				{[}-4.0,-5.0) & 6 & 3  & {[}4.0,5.0) & 39 & 26  \\
				{[}-3.0,-4.0) & 30 & 15 & {[}3.0,4.0) & 114 & 76  \\ \rowcolor{gray!20}
				{[}-2.0,-3.0) & 110 & 48 & {[}2.0,3.0) & 266 & 175  \\
				{[}-1.0,-2.0) & 298 & 157 & {[}1.0,2.0) & 455 & 307  \\ \rowcolor{gray!20}
				{[}0.0,-1.0) & 359 & 203 & {[}0.0,1.0) & 430 & 345  \\ \midrule
				{\bf Total} & 807 & \multicolumn{1}{>{\columncolor{mygray}}c|}{427} &  & 1323 & \multicolumn{1}{>{\columncolor{mygray}}c}{941}  \\ \bottomrule
			\end{tabular}}
\end{table}
We inferred 1,952 (91.64\%=1952/2130) classification rules in total based on the inferred features.
However, not all of them can be used for evasion attacks because some of them are related to appearance or functionalities or URLs.
Table~\ref{table:rule decryption} shows the 1,327 classification rules that can be used for our evasion attacks, in which 427 classification rules
are negative classification rules  that can be used for addition, and 941 classification rules are positive classification rules that can be used for deletion.
We remark that it is unnecessary to fully resolve the features of these inferred classification rules,
as we can only know partial features of some classification rules.
Such classification rules can only be deleted during evasion attacking.
Therefore, classification rule deletion is considered to be more effective than classification rule addition.

\begin{table}[t]\footnotesize
		\caption{Performance of our attacks in white-, grey- and black-box scenarios}
		\label{table:performance of attacks}\centering
	\scalebox{0.95}{
	\begin{tabular}{c|c|c|c|c}
				\toprule
				\multirow{2}{*}{\textbf{Score}} &  \textbf{\#Phishing}  & {\bf \#Succeeded } & {\bf Mutation} & {\bf \#Mutated}  \\
                                                & \textbf{Websites}  & {\bf \#Websites} & {\bf Time (ms)} & {\bf Features/Rules}\\ \toprule
                \multicolumn{5}{c}{Performance of White-box Attacks}\\\midrule
				\rowcolor{gray!20}
				1 & 1 & 1 & 572.00 & 5.00/30.00 \\
				{[}0.9,1.0) & 131 & 131 & 91.00 & 1.65/9.81  \\
				\rowcolor{gray!20}
				{[}0.8,0.9) & 59 & 59 & 49.59 & 1.10/4.51  \\
				{[}0.7,0.8) & 14 & 14 & 48.71 & 1.00/3.93  \\
				\rowcolor{gray!20}
				{[}0.6,0.7) & 68 & 68 & 26.35 & 1.00/4.32  \\
				{[}0.5,0.6) & 32 & 32 & 32.22 & 1.00/2.31  \\ \midrule
				\textbf{Total} & \textbf{305} & \textbf{305} & - & -  \\ \bottomrule

                \multicolumn{5}{c}{Performance of Grey-box Attacks}\\\midrule
                \rowcolor{gray!20}
				1 & 1 & 1 & 966.00 & 14.00/48.00  \\
				{[}0.9,1.0) &  131 &  131& 165.76 & 6.91/19.47  \\
				\rowcolor{gray!20}
				{[}0.8,0.9) &  59 &  59& 121.63 & 6.22/15.61  \\
				{[}0.7,0.8) &  14 &  14 & 194.86 & 5.50/12.43  \\
				\rowcolor{gray!20}
				{[}0.6,0.7) & 68 & 68 & 139.25 & 5.01/11.87  \\
				{[}0.5,0.6) & 32 &32 & 106.38 & 3.84/7.13 \\ \midrule
				
				\textbf{Total} & \textbf{305} & \textbf{305} & - & - \\ \bottomrule
                \multicolumn{5}{c}{Performance of Black-box Attacks (node modification only)}\\\midrule
		
			    1 & 1  & 1 & 606.32 & 24.00/50.00 \\
				{[}0.9,1.0) & 131 &  124 & 244.96 & 7.54/19.76 \\
				\rowcolor{gray!20}
				{[}0.8,0.9) & 59  & 59 & 206.34 & 3.15/9.76 \\
				{[}0.7,0.8) & 14 & 14 & 291.00 & 1.79/7.93 \\
				\rowcolor{gray!20}
				{[}0.6,0.7) & 68  & 68 & 231.97 & 2.21/8.14 \\
				{[}0.5,0.6) & 32 & 32 & 219.94 & 5.32/10.03 \\ \midrule				
				\textbf{Total} & \textbf{305} & \textbf{298} & - & - \\ \bottomrule
			\end{tabular}}
 \end{table}	

	\subsection{\bf Evaluation of Evasion Attacks}\label{sec:evaluation-evasion-attack}
	\subsubsection{Effectiveness and Efficiency}
To evaluate effectiveness and efficiency of our evasion attacks, we respectively conduct white-, grey- and black-box attacks against the classifier \texttt{GPPF} using 305 phishing websites as seeds.
The results of our proposed evasion attacks are shown in Table~\ref{table:performance of attacks}.
Column ``Score'' shows the decision  score range, with corresponding number of phishing websites in
Column ``\#Phishing Websites''.
Columns ``\#Succeeded Websites'' indicate the number of crafted adversarial websites.
Columns ``Mutated Time'' indicate the average time spent on crafting the adversarial websites.
Columns ``\#Mutated Features/Rules'' show the average number of mutation operations (i.e., deletion and addition) for crafting the adversarial websites,
which also indicate query times to the classifier.
Remark that the number of mutated classification rules is always larger than that of features,
as when one feature is mutated, classification rules that rely on this feature are also mutated.

	

    From the results in Table~\ref{table:performance of attacks},
	we can see that our white- and grey-box attacks achieve 100\% attack success rate,
and our black-box attack successfully crafts 298 adversarial samples by only modifying DOM nodes for 305 seeds.
    (Note that in the black-box scenario, we first modify nodes and then add nodes,
    while node modification and addition operations may be mixed during mutation in white- and grey-box scenarios.
    Although features are deleted first in grey-box attacks, the feature deletion is achieved by node modification and addition.)

	The black-box attack failed to craft adversarial samples for 7 websites (cf. Table~\ref{table:remain 7 phishing pages}) by only modifying nodes.
   Therefore, we continue mutating them by adding nodes.
Table~\ref{table:remain 7 phishing pages} shows the results,
which target five domains.
Column ``Score after modification'' gives the decision score after node modification.
Column ``\#Addition Operations'' gives the number of node addition operations.
Column ``Mutate Time'' gives the time of node addition operations.




\begin{table}[t]\scriptsize
	 \caption{7 phishing websites that fail to evade {\tt GPPF} by only modifying nodes in black-box scenario}
	\label{table:remain 7 phishing pages}\setlength{\tabcolsep}{3pt}
	 	\scalebox{0.95}{
		\begin{tabular}{cccc}   \hline
			\textbf{Target Domain} & \begin{tabular}[c]{@{}c@{}}\textbf{Score after}\\ \textbf{Modification}\end{tabular} & \begin{tabular}[c]{@{}c@{}}\textbf{\#Addition}\\ \textbf{Operations}\end{tabular} &\begin{tabular}[c]{@{}c@{}}\textbf{Mutate}\\ \textbf{Time (ms)}\end{tabular}\\ \toprule
			sso.godaddy.com/login & 0.81 & 543  & 62.87\\
			passport.alibaba.com/icbu\_login.htm & 0.83 & 543  & 56.12\\
			\rowcolor{gray!20}
			store.cpanel.net/login/  & 0.84 & 543 & 60.10\\
			store.cpanel.net/login/  & 0.84 & 543 & 57.61\\
			\rowcolor{gray!20}
			store.cpanel.net/login/  & 0.84 & 543 & 58.27\\
			passport.alibaba.com/icbu\_login.htm  & 0.90 & 543 & 60.76\\
			\rowcolor{gray!20}
			passport.alibaba.com/icbu\_login.htm  & 0.93 & 1032 & 299.54\\  \bottomrule
		\end{tabular}}
 \end{table}

 From the results in Table~\ref{table:remain 7 phishing pages}, we can observe that the adversarial samples of all the 7 websites
  can be crafted by node addition, using at least 543 steps.
  After a deep analysis of these added nodes,  we found that only one or two classification rules are changed.
  This demonstrates that node modification is more efficient than node addition in  black-box scenario.
	The most difficult website is the last one whose initial decision  score is $0.99$.
After modifying all modifiable nodes, its decision  score becomes $0.93$, which is still larger than the threshold $0.5$.
It takes 1032 times of addition operations to reduce its decision score below $0.5$.
Table~\ref{table:rules can't be deleted in case2} show the details of classification rules that are hit by the website, but cannot be deleted.
Column ``Features'' represents the features of each classification rule.
Although we delete many positive classification rules from the webiste, the weights of the remaining classification rules hit by the websites are still larger enough so that
  the decision score is still larger than $0.5$.

\begin{table}[t]\footnotesize
	\centering\setlength{\tabcolsep}{3pt}
	\caption{Classification rules that cannot be deleted in the real case}
	\vspace{-2mm}
	\label{table:rules can't be deleted in case2}	
	\begin{tabular}{cccc}
			\toprule
		{\bf Rule} & {\bf Features} & {\bf Rule Weight} & {\bf }\\ \hline  
		1 & \begin{tabular}[c]{@{}c@{}}PageNumScriptTags\textgreater{}6, PageExternalLinksFreq\end{tabular} & 0.55 \\ \midrule
		2 & UrlPathToken=$*$ & 2.23 \\ \midrule   
		3 & \begin{tabular}[c]{@{}c@{}c@{}}PageNumScriptTags\textgreater{}1,PageNumScriptTags\textgreater{}6 \\
		   PageExternalLinksFreq	
		\end{tabular} & 0.52 \\ \midrule
		4 & \begin{tabular}[c]{@{}c@{}}UrlPathToken=$*$, PageHasForms\end{tabular} & 0.21 \\ \hline    
		5 & \begin{tabular}[c]{@{}c@{}}PageNumScriptTags\textgreater{}6, UrlPathToken=$*$\end{tabular} & 0.46 \\ \midrule
	     6 & \begin{tabular}[c]{@{}c@{}}PageNumScriptTags\textgreater{}1, UrlPathToken=$*$\end{tabular} & 1.17 \\ \midrule    
		7 & \begin{tabular}[c]{@{}c@{}}PageNumScriptTags\textgreater{}1, UrlPathToken=$*$\end{tabular} & 0.43 \\ \midrule
		8 & PageImgOtherDomainFreq & 0.34 \\ \hline    
		9 & \begin{tabular}[c]{@{}c@{}}PageNumScriptTags\textgreater{}1, PageExternalLinksFreq\end{tabular} & 1.48 \\ \bottomrule
	\end{tabular}
\end{table}
	
%
	

\subsubsection{Visual Impact}
To evaluate the visual impact of the mutated webpages,
we capture screenshots of all webpages of phishing and crafted adversarial websites
and compare each phishing screenshot with its adversarial ones using mean-squared error (MSE)~\cite{wb09}
and mean-absolute error (MAE)~\cite{chai2014root} criterion,
two common distortion measures of images.
All the values of MAE and MSE are $0$, which demonstrates that
our attacks preserve appearance.
We discuss two cases of visual impact which respectively target Yahoo
and PayPal.

\begin{figure}[t]
	\centering
	\begin{subfigure}[t]{0.25\textwidth}
		\centering
		\includegraphics[width=1\textwidth]{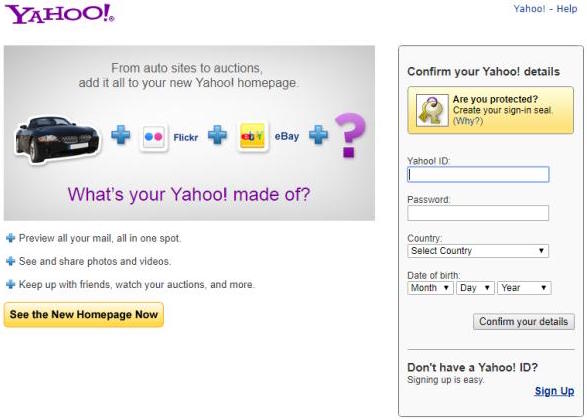}
		\caption{Original phishing webpage}
		\label{fig:yahoo-origin}
	\end{subfigure}%
	\begin{subfigure}[t]{0.25\textwidth}
		\centering
		\includegraphics[width=1\textwidth]{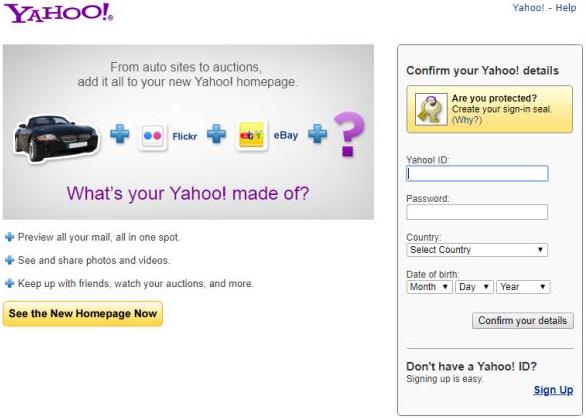}
		\caption{Adversarial sample crafted in grey-box scenario}
		\label{fig:yahoo-mutated}
	\end{subfigure}
	\caption{Screenshots of two webpages targeting yahoo.}
	\label{fig:visual-comp-yahoo}
\end{figure}

Fig.~\ref{fig:visual-comp-yahoo} shows the screenshots of the original phishing website  targeting Yahoo (Fig.~\ref{fig:yahoo-origin}) and an adversarial sample (Fig.~\ref{fig:yahoo-mutated}),
crafted in grey-box scenario. Our tool added seventeen {\tt PageTerm} features, one  {\tt PageImgOtherDomainFreq} feature, and one {\tt PageHasPswdInputs} feature,
deleted one {\tt PageExternalLinkFreq} feature and two {\tt PageTerm} features.
Adding {\tt PageImgOtherDomainFreq} and deleting {\tt PageExternalLinkFreq} features
are achieved by adding 3,321 {\tt <a>} nodes with internal href and  554 {\tt <img>} nodes with external href.
The values of both MAE and MSE are 0.

\begin{figure}[t]
	\centering
	\begin{subfigure}[t]{0.25\textwidth}
		\centering
		\includegraphics[width=\textwidth]{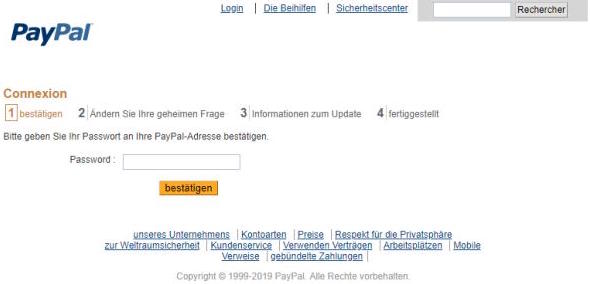}
		\caption{Original phishing webpage}
		\label{fig:paypal-origin}
	\end{subfigure}%
	\begin{subfigure}[t]{0.25\textwidth}
		\centering
		\includegraphics[width=\textwidth]{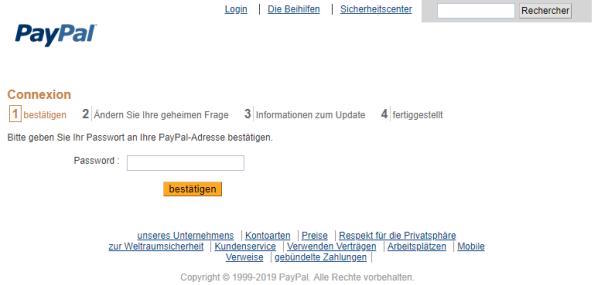}
		\caption{Adversarial sample crafted in black-box scenario}
		\label{fig:paypal-mutated}
	\end{subfigure}
	\caption{Screenshots of two webpages targeting PayPal.}
	\label{fig:visual-comp-paypal}
\end{figure}
Fig.~\ref{fig:visual-comp-paypal} shows the screenshots of a phishing website  targeting PayPal (Fig.~\ref{fig:paypal-origin})
and an adversarial sample (Fig.~\ref{fig:paypal-mutated}),
crafted in black-box scenario.
Our tool deleted three{\tt PageTerm} features, one {\tt PageHasPswInputs} feature, and one {\tt PageLinkDomain} feature.
The values of both MAE and MSE are 0.


	\begin{table}[t]
		\centering
		\footnotesize
		\caption{Transferability of crafted samples}
		\label{table:transferability of crafted samples}
		\begin{tabular}{ccc} \toprule
			\textbf{Attack} & \textbf{\#Detected/\#Crafted} & \textbf{Detection Ratio} \\ \midrule
			White-box & 32/32 & 100\% \\
			Grey-box & 16/32 & 50\% \\
			Black-box & 6/32 & 18.75\% \\ \bottomrule
		\end{tabular}
	\end{table}

\smallskip
\noindent{\bf Transferability Attack.}
%
%
%
%
As aforementioned, we evaluate the transferability attack on BitDefender
{TrafficLight}. Table~\ref{table:transferability of crafted samples} shows the results of the transferability attack on BitDefender
{TrafficLight}.
We can observe that adversarial samples crafted by the black- and grey-box attacks achieved 81.25\% and 50\% transferability attack rate.
In contrary, BitDefender {TrafficLight} detected all the adversarial samples crafted by the white-box attacks.
This is because that our white-box attack only made minor changes that are
very specific to features/rules of \texttt{GPPF},
whereas black-box attack made more changes independent of \texttt{GPPF},
hence, is more robust than others for transferability attacks.

As aforementioned, BitDefender TrafficLight is
completely black-box and part of its processing is done in
the cloud with some intelligent engines, therefore it is impossible to find the reason why
adversarial examples crafted by our attacks on \texttt{GPPF}
are still effective against BitDefender TrafficLight.
One possible reason is that BitDefender TrafficLight may leverage similar features as \texttt{GPPF} to decide whether a website is phishing or not.

\subsection{Evaluation of Defense Methods}\label{sec:eval-defense}

 \begin{figure}[t]
 	\centering
	\includegraphics[width=0.45\textwidth]{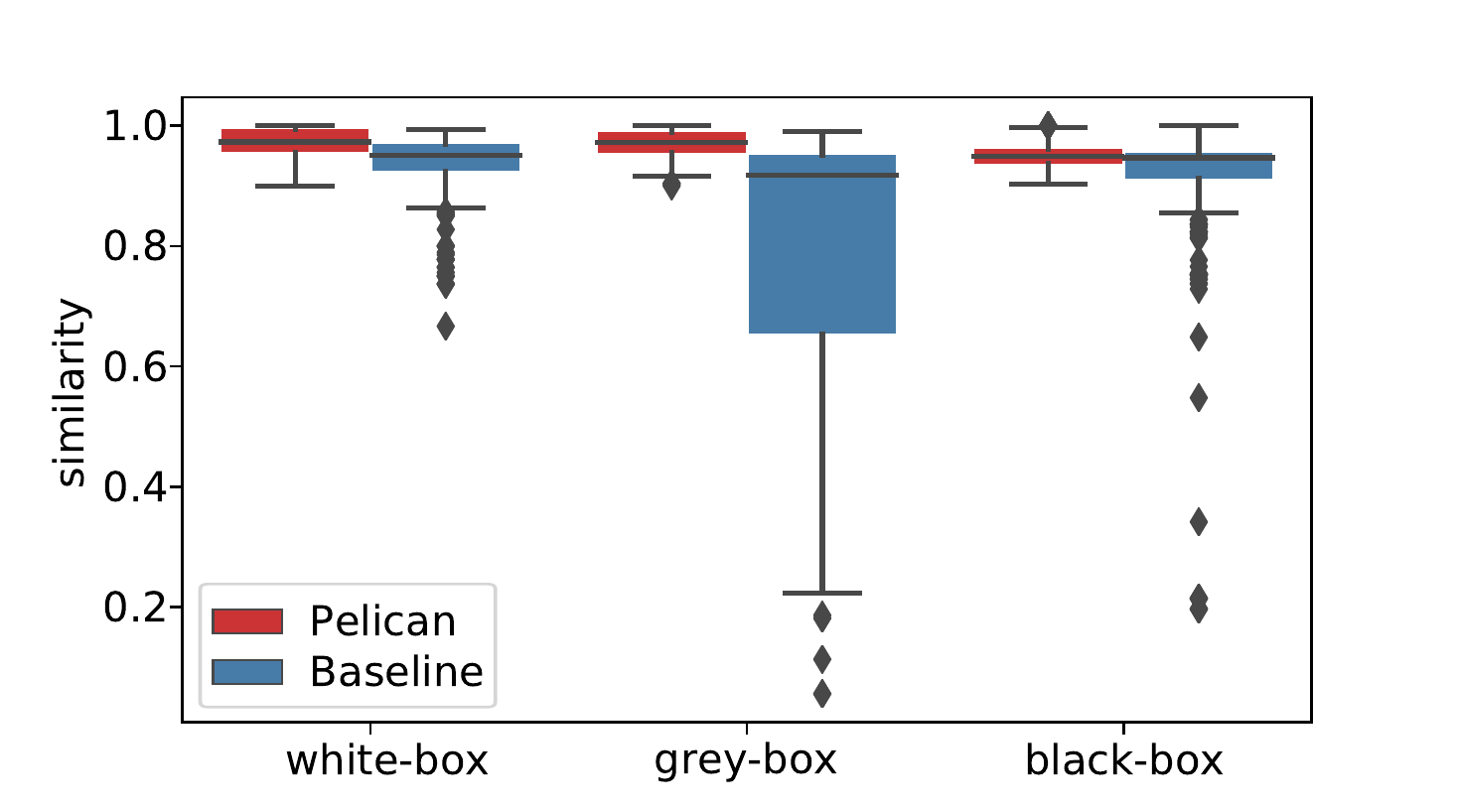}
 	\captionof{figure}{Results of similarity comparison.}\label{fig:defense}
 \end{figure}

\subsubsection{\bf Evaluation of Similarity-based Evasion Detection}
We evaluate both DOM tree similarity method and \tool
against three evasion attacks.
For each method, we compare the similarity between
the adversarial sample and its original phishing website.
We use the 305 original phishing websites and their corresponding adversarial
samples as the subjects.
Fig.~\ref{fig:defense} shows the results of similarity comparison using two methods under three attacks.
 We can observe that both DOM tree similarity method and \tool achieve high similarity (i.e., over {90\%}) on most of the
 adversarial samples crafted by white- and black-box attacks.
 Surprisingly, on the adversarial websites crafted by grey-box attacks,
 \tool significantly outperforms the DOM tree similarity method.
 After a further investigation of these websites,
 we find that there are $4$ frequency-related features in \texttt{GPPF},
 adding or deleting these features needs to mutate a lot of nodes in order to reduce/increase the frequency of some elements.
 Hence, mutation of these features adds lots of invisible nodes,
 which reduces DOM tree similarity by increasing the number of nodes. 

 For example, a phishing website with domain ``{login.microsoftonline.com.best10reviews.com}''
 has decision score 0.92. It hits a classification rule with weight 2.00 in the classifier, the classification rule consists of two features: ``{PageSecureLinksFreq}'' and ``{PageHasPswdInputs}''.
 ``{PageSecureLinksFreq}'' is a frequency-related feature, computed as ${\tt \frac{\#secure\_links}{\#total\_links}}$. Currently, it equals to $1$,
 indicating that all the {``href''} links in the website are secure (i.e., ``https").
 In order to delete this feature, our grey-box attack adds
 lots of insecure (i.e., ``http") and invisible {``href''} links.
 Consequently, the DOM tree similarity is $46.78\%$, thus can bypass the DOM tree similarity based detection.
 However, such frequency-related features have no effect on \tool,
 as \tool computes the similarity of the number of nodes in the phishing website, instead of that in the crafted sample.
 Thus, such websites can be detected by \tool.

\subsubsection{\bf Evaluation of Subset Classification Rule Pruning}\label{sec:expsubsetruleimprove}
We found that \texttt{GPPF} has at least 150 addable negative subset classification rules.
Therefore, we conduct experiments on {\tt GPPF} without subset classification rules to investigate the impact on attack cost
and accuracy.
To remove subset classification rules in {\tt GPPF}, we set the values of such classification rules
to $0$, as it is impossible for us to retrain {\tt GPPF}.

\smallskip
\noindent{\bf Impact on attack cost.}
{We automatically create 50 personalized phishing websites (10 websites in each decision score range)
as our subjects.
Inspired by the 7 phishing websites in Table~\ref{table:remain 7 phishing pages}
on which feature deletion fails to craft adversarial examples
in black-box scenario,
we first collect static webpages from legitimate websites and change the action field of HTML forms in these websites for phishing,
then delete all deletable features,
and finally add some undeletable features to make their decision scores within the desired range.}

{We compare the attack costs with and without using subset classification rules in  black-box scenario.
Table~\ref{table:result of reducing inclusion rules} shows the results.
We can observe that the attack requires more operations to craft an adversarial sample in black-box scenario without using subset classification rules,
hence, conclude that pruning subset classification rules can increase the attack cost.} 

\smallskip
\noindent{\bf Impact on accuracy.}
We check all 305 original phishing websites and
15,000 legitimate website using {\tt GPPF} without subset classification rules.
Although, there are 2,017 legitimate and 232 phishing websites whose decision decision  scores changed after disabling subset classification rules, but the changes are small so that none of them are misclassified by the classifier.

\begin{table}[t]
	\footnotesize\setlength{\tabcolsep}{3pt}
	\centering
	\caption{Attack cost comparison with/without subset classification rules}
	\label{table:result of reducing inclusion rules}
	{\begin{tabular}{cccc}
		\toprule
		{\bf Score}       & {\bf \#Websites} & \begin{tabular}[c]{@{}c@{}}{\bf \#Operations} {\bf (With)}  \end{tabular} & \begin{tabular}[c]{@{}c@{}}{\bf \#Operations} {\bf (Without)}\end{tabular} \\ \midrule  \rowcolor{gray!20}
		{[}0.9,1.0) & 10     & 1331.7                                                                      & 3015.0                                                                        \\
		{[}0.8,0.9) & 10     & 643.4                                                                       & 1028.8                                                                         \\   \rowcolor{gray!20}
		{[}0.7,0.8) & 10     & 418.2                                                                       & 967.0                                                                        \\
		{[}0.6,0.7) & 10     & 273.5                                                                       & 353.8                                                                          \\    \rowcolor{gray!20}
		{[}0.5,0.6) & 10     & 212.3                                                                     & 221.0                                                                          \\ \bottomrule
	\end{tabular}}
\end{table}

\begin{table}[t]
	\footnotesize\setlength{\tabcolsep}{3pt}
	\centering
	\caption{Single classification rules in \texttt{GPPF}}
	\label{table:single_rules_in_gppf}
	\scalebox{0.9}{
	\begin{tabular}{ccccc}
		\toprule
		{\bf Feature} & {\bf \#Deletable} & {\bf Total  Weight}  & {\bf \#Addable} & {\bf Total  Weight}  \\ \midrule
		PageTerm=$*$ & 23 & 43.42 & 14 & -22.94 \\
		PageLinkDomain=$*$ & 16 & 39.96 & 20 & -19.89 \\ \midrule
		Total & 39 & 83.38 & 34 & -42.83 \\ \bottomrule
	\end{tabular}}
\end{table}
\subsubsection{\bf Evaluation of Single Classification Rule Pruning}\label{sec:expsingleruleimprove}
We found that \texttt{GPPF} has at least 34 addable and 39 deletable single classification rules, as shown in Table~\ref{table:single_rules_in_gppf}.
Therefore, we conduct experiments on \texttt{GPPF}.
%


\smallskip
\noindent{\bf Impact on attack cost.}
{To understand the impact on attack cost, we conduct experiments in white-box and grey-box scenarios by only using single classification rules for mutation.
We achieved 100\% attack success rate using only these single classification rules.
An adversarial sample is crafted in less than 1ms for the 305 phishing websites.
Therefore, we conclude that single classification rules have a potentially severe hazard to classifiers,
and should be pruned from the classification rules in order to enhance the robustness.}

\smallskip
\noindent{\bf Impact on accuracy.}
{To understand the impact on accuracy, we set the values of all deletable or addable single classification rules
to $0$ for all websites.
We check all 305 original phishing websites and
15,000 legitimate websites using {\tt GPPF}.
Although, there are 60 legitimate and 29 phishing websites whose decision score changed after
disabling single classification rules, none of them is misclassified by {\tt GPPF}.}

\section{Discussion}\label{sec:discussion}
In this section, we discuss the limitations of our study and other suggestions for enhancement of classifiers.



\smallskip
\noindent{\bf Limitations.}
Our attacks are nondestructive, i.e., do not change appearance and functionalities.
This provides a chance to defenders to add and enhance features by adding appearance-related features.
As shown in our experiments, if the website contains many undeletable features and its decision  score is very large,
it is relatively much more difficult to craft adversarial samples in black-box scenario.
The node addition operation in our black-box attack depends on the nodes of the legitimate websites we collected.
Thus, it is likely to happen that the collected nodes from the legitimate websites are not comprehensive enough,
causing failure in black-box scenario, even though this situation did not happen in our experiments.

Our attack assumes that the target classifier
outputs the decision score for each input test.
This is possible for client-side classifiers by leveraging reverse
engineering techniques.
For classifiers that cannot output decision score,
we could launch attacks in the white-box scenario
by deleting all the positive classification rules
and adding many negative classification rules in one step.
However, in the grey- or black-box scenarios,
our mutation cannot be guided by the decision scores.
One possible solution is to delete DOM nodes that
occur frequently in phishing websites but rarely in legitimate websites,
and add DOM nodes that occur
frequently in legitimate websites but rarely in phishing websites.
We leave them as future work.

Furthermore, it might be able to improve resilience of classifiers
by proposing new classification rules or conducting adversarial retraining using crafted adversarial samples, following
disclosure of our findings.
But, in both cases, our attacks could achieve node manipulation
via more complex JavaScript code and {Shadow DOM~\cite{HAJSS14}}, which are widely used in legitimate websites,
but very difficult to analyze statically. This might increase attack cost.
On the other hand, attackers can aggressively add or modify nodes to evade our defense method, however, by sacrificing the appearance or functionality of the websites, making them less convincing.
Moreover, our approach raises the difficulty bar for evasion attacks.

For the defense method, \tool is weak for the scenario where attackers generate the crafted adversarial samples directly rather than modifying the original one detected by classifiers.
\tool stores the last $k$ or the ones queried in the last $h$ hour,
rendering the tool inefficient when $k$ or $h$ are sufficiently large.
This could be avoided by leveraging cloud infrastructure, as done in Monarch~\cite{ThomasGMPS11}.







\smallskip
\noindent{\bf Other Suggestions for Classification Rule Enhancement.}
According to the characteristics of different features we analyzed,
we give advice to enhance the robustness of the classifier \texttt{GPPF}.
In node modification operation, the order of attack cost is: 
\begin{center}
\small	{\tt Term-related feature} $<$ {\tt Dom-related feature}.
\end{center}
In node addition operation, the order is: 
\begin{center}
\small	{\tt Term-related feature} $=$ {\tt Dom-related feature}.
\end{center}
According to our experimental results,
term-related features are very weak, as it can be easily added or deleted,
while Dom-related features can be added, but only some of them can be deleted.
Therefore,
(1) under attacks using node addition,
it is better to combine term- and Dom-related features with other features which cannot be added easily in negative classification rules with low weights;
(2) under attacks using node modification,
it is better to prune term-related features and combine Dom-related features with
other features that cannot be deleted easily in positive classification rules with high weights.
In general, defenders should minimize the use of term-related features
and enhance existing classification rules with
appearance-related features, making these classification rules difficult to be deleted or added without changing the functionality and appearance of the websites.

\section{Related Work}\label{sec:relatedwork}

\noindent{\bf Phishing website detection.}
 {   Phishing website detection has been studied in
    several work, such as blacklist~\cite{ludl2007effectiveness, oest2019phishfarm}, similarity-based~\cite{LiuDHF06,medvet2008visual,AfrozG11,fu2006detecting,rosiello2007layout},
    heuristic-based~\cite{teraguchi2004client,kirda2006protecting}, and ML-based ~\cite{zhang2007cantina,whittaker2010large, corona2017deltaphish, ubing2019phishing} anti-phishing solutions.
    The blacklist approach maintains a list of URLs of known phishing websites,
    but it is only able to detect phishing websites whose URLs are in the list.
    Similarity-based approaches compare similarity between authentic and phishing webpages,
    differing in terms of: features extracted to identify similarity and the matching algorithm used.
   Heuristic-based approaches design specific detection rules based on common characteristics of phishing websites.
   Similarity- and heuristic-based approaches are able to detect
    new phishing websites, although not greatly improve the accuracy of other existing approaches~\cite{MAGSSA17,VMA16}.
    ML-based approaches train a classification model using features/rules from known phishing websites and  authentic websites,
    differing in terms of: features for training and machine learning algorithm used.
   They are scalable and accurate.
   We refer to \cite{VMA16} for a survey of web phishing detection approaches.}

   However, blacklist-based classifiers are shown vulnerable to evasion attacks by Oest et al.~\cite{oest2019phishfarm}.
    In this work, we showed that ML-based classifiers are also vulnerable to evasion attacks even in black-box scenario.
    We conjecture that classifiers based on other approaches are also vulnerable to evasion attacks, in particular in white- and grey-box scenarios,
    for which the adversary can gain details of classifiers.

 The similarity-based approaches~\cite{LiuDHF06,medvet2008visual,AfrozG11,fu2006detecting,rosiello2007layout} are closely related to our defense
method \tool.  The main differences are:
(1) \tool aims at detecting evasion attacks rather than general phishing websites;
(2)  \tool compares the similarity between an unknown one and recently detected phishing webpages rather than
the similarity between an unknown one and authentic webpages;
and (3) we compute similarity layer-by-layer and only need to record recently detected phishing webpages due to (1), hence
is both time and space efficient.

	
\smallskip
\noindent{\bf Attacks on classifiers.}
{Evasion attacks} and {poisoning attacks} are two main types of attacks on ML-based classifiers.
Evasion attack aims to evade classifiers by crafting adversarial samples using various mechanisms
and have been exhibited in different scenarios such as
malicious PDF files~\cite{maiorca2013looking,SL14,xu2016automatically},
malware~\cite{grosse2017adversarial, demontis2017yes, chen2018automated}, malicious websites~\cite{xu2014evasion}, spam emails~\cite{lowd2005good}, and speaker recognition systems~\cite{chen2019real}.
However, domain-specific characteristics that usually pose new challenges should be token into account in different scenarios.
For example, adversarial images usually minimize distortion when adding pixel-level noise~\cite{grosse2017adversarial,SBR18},
and adversarial (PDF) malware usually need to keep original malicious functionalities~\cite{chen2018automated,xu2016automatically} \emph{only}.
However, adversarial phishing websites usually have to preserve both appearance and functionalities which
arise new challenges compared with other scenarios.
These new challenges potentially make this research area less noticed and more difficult.
{For instance, although the method proposed by Xu et al.~\cite{xu2016automatically} is generic for crafting adversarial samples.
This method leverages an evolutionary algorithm to mutate PDF malware and relies an oracle to determine if a generated variant
preserves maliciousness. It is non-trivial to be directly adapted into our setting. }
In addition, in recent years, attacks on deep neutral network based-classifiers are also widely studied (e.g., \cite{SBBR19,grosse2017adversarial, carlini2017adversarial,hu2018black}), which have the similar theory by leveraging adversarial images in machine learning.
In summary, all these existing studies consider domain-specific scenarios with many differences compared with ours in phishing website detection
which owns new
challenges.

\begin{figure}[t]
\begin{subfigure}[t]{.25\textwidth}
		\centering
		\includegraphics[width=.9\textwidth]{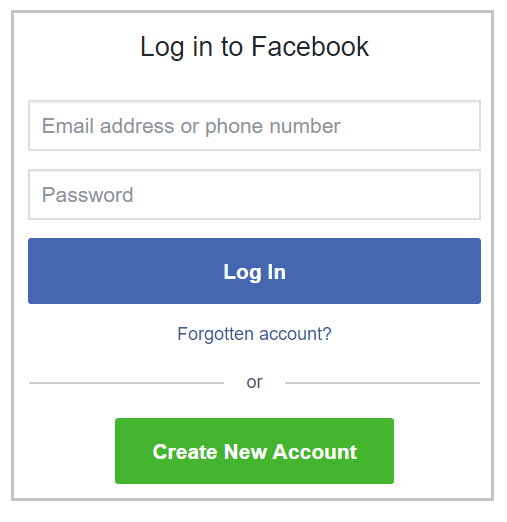}
		\caption{Original one}
		\label{fig:facebook-old}
	\end{subfigure}%
\begin{subfigure}[t]{0.25\textwidth}
		\centering
		\includegraphics[width=.9\textwidth]{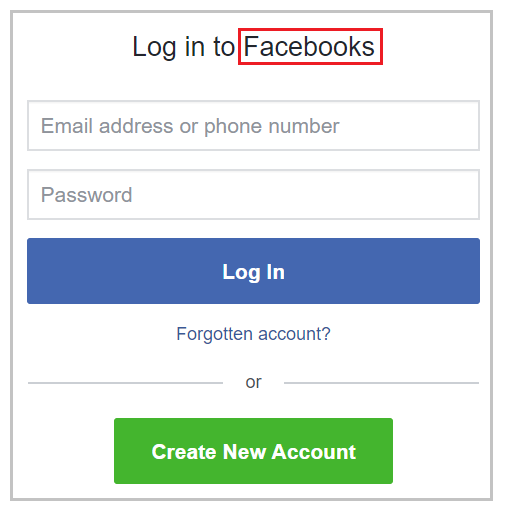}
		\caption{Adversarial one crafted by~\cite{liang2016cracking}}
		\label{fig:facebook-new}
	\end{subfigure}
	\caption{Visual comparison. }
	\label{fig:comp} 	
 \end{figure}

\smallskip
\noindent{\bf Comparison over the attack in~\cite{liang2016cracking}}.
The most closely related work is Liang et al.~\cite{liang2016cracking}, which proposes
collision attack to infer features and evasion attacks on phishing website classifier {\tt GPPF}.
There are seven key differences.
\begin{enumerate}
\item The collision attack in Liang et al.~\cite{liang2016cracking} uses alphabets
and full-text corpora with seven languages to infer features, while our collision attack infers
features by leveraging data from phishing and legitimate websites.
Experimental results demonstrate that our collision attack is more effective and efficient than the one of \cite{liang2016cracking} (cf. Table~\ref{table:feature decryption}.).

  \item Our evasion attacks do not destruct the appearance and functionalities of webpages,
    whereas the method in \cite{liang2016cracking} may do,
    making them less effective for web phishing.
    For example, Fig.~\ref{fig:comp} shows the two webpages:
    the original one (Fig.~\ref{fig:facebook-old}) and the adversarial one (Fig.~\ref{fig:facebook-new}) crafted by~\cite{liang2016cracking},
which is obviously illegal.

  \item Our white-box attack leverages a greedy algorithm to maximally decreasing the decision  score at each mutation step, while \cite{liang2016cracking} only either adds positive classification rule or deletes negative classification rules, hence less efficient than ours.

  \item The more important and practical yet more challenging grey- and black-box attacks are not considered by Liang et al.~\cite{liang2016cracking},
while we propose effective and efficient grey- and black-box attacks by leveraging a greedy algorithm.

  \item We demonstrate transferability attack on the industrial
tool Bitdefender TrafficLight which is completely black-box. Our attack achieved up to 81.25\% success rate using black-box attacks.
Whereas, Liang et al.~\cite{liang2016cracking} do not consider either the transferability attack or other ML-based classifiers.

  \item We propose defense methods against evasion attacks which are effective to identify adversarial samples. However this is not considered by Liang et al.~\cite{liang2016cracking}.
  \item We also propose two classification rule selection strategies that are able to enhance the robustness
of the classifiers and increase the attack cost, which are not considered by Liang et al.~\cite{liang2016cracking}.
\end{enumerate}

\textit{Poisoning attacks} can fully manipulate almost all the training data and features, and finally train a bad classifier without abilities for classification~\cite{biggio2012poisoning}.
Spam detection and intrusion detection are always vulnerable to poisoning attacks~\cite{nelson2008exploiting, biggio2014security}.
However, we believe it is very difficult to launch a poisoning attack on practical phishing website classifiers (e.g., \texttt{GPPF}), as the training data of phishing website classifiers
is huge and its source is also diverse.

\smallskip
	\noindent{\bf Defenses for classifiers.}
It is not surprising that many defense solutions have been proposed to enhance the robustness of ML-based classifiers.
The protection of training and features can help to train a more robust classifier and increase the difficulty of evasion~\cite{saeys2008robust,biggio2010multiple,smutz2016tree}.
For example, Saeys et al. ~\cite{saeys2008robust} use ensemble feature selection techniques and demonstrate that multiple feature selection methods can receive more robust results.
Biggio et al.~\cite{biggio2010multiple} use multiple classifiers to gain a better results in adversarial environments.
GoodFellow et al.~\cite{goodfellow2014explaining} re-train on adversarial samples to improve the robustness of classifiers, and highlight that ensemble methods have some limitations for evasion attacks.
Zhang et al.~\cite{zhang2016adversarial} use adversarial feature selection to increase the robustness of classifiers.
In the area of malware, Chen et al.~\cite{chen2018automated, chen2016towards} propose a similarity-based approach for filtering Android crafted samples by leveraging the most benign and malicious samples.
In addition, many recent studies focus on identifying adversarial samples in deep neural networks~\cite{liang2017detecting,xu2017feature}.
These solutions are all preventive in advance.
However, all these studies are tailored to different setting than ours.
It is difficult to apply these solutions to enhancing the robustness of phishing website classifiers directly due to the domain-specific characteristics.

As for the client-side classifiers, the effective solution is to increase the complexity of reverse engineering by using more advanced obfuscation techniques. The less information an attacker gets from a classifier, the safer the classifier is.
In terms of the defense solutions for phishing website classifiers with evasion attacks, to our best knowledge, there is no existing studies that focus on it.

\section{Conclusion and Future Work}\label{sec:conclusion}
In this work, we conducted the first systematic study on evasion attacks of ML-based web phishing classifiers
and revealed their weaknesses.
Specifically, we proposed a novel sample-based collision attack
which is able to effectively and efficiently infer almost all the hashed features and classification rules of the classifier \texttt{GPPF}.
It is more effective and efficient than the recent collision attack technique~\cite{liang2016cracking}.
We also proposed a class of advanced evasion attacks for three attack scenarios (i.e., white-, grey-, and black-box) by leveraging mutation
techniques and the first defense method against evasion attacks.
We further presented two classification rule selection strategies to enhance the robustness of the classifiers and increase the attack cost.
Experimental results demonstrated that our attacks can achieve 100\% attack success rate in less than one second per seed on \texttt{GPPF},
and 81.25\% transferability attack rate on \texttt{TrafficLight}.
Our defense method is effective in detecting crafted adversarial websites, hence can be used to
detect evasion attacks and enhance the robustness of classifiers.
Our study shows that sophisticated phishing attack is still a severe cyber security vector.

Currently, our attack mutates seeds by leveraging a greedy algorithm.
In future, we plan to investigate attacks by leveraging genetic algorithms, which have been used
to fool PDF malware classifiers~\cite{xu2016automatically},
face recognition systems~\cite{SBBR16} and
image recognition systems~\cite{ASCZHS19}.
It is interesting to understand which algorithm is more effective and efficiency.
Another future direction is to investigate attacks
against ML-based classifiers  that only outputs
the decision result instead of decision score for each input test.


\bibliographystyle{IEEEtran}
\bibliography{jrnl}

\begin{thebibliography}{10}
\providecommand{\url}[1]{#1}
\csname url@samestyle\endcsname
\providecommand{\newblock}{\relax}
\providecommand{\bibinfo}[2]{#2}
\providecommand{\BIBentrySTDinterwordspacing}{\spaceskip=0pt\relax}
\providecommand{\BIBentryALTinterwordstretchfactor}{4}
\providecommand{\BIBentryALTinterwordspacing}{\spaceskip=\fontdimen2\font plus
\BIBentryALTinterwordstretchfactor\fontdimen3\font minus
  \fontdimen4\font\relax}
\providecommand{\BIBforeignlanguage}[2]{{%
\expandafter\ifx\csname l@#1\endcsname\relax
\typeout{** WARNING: IEEEtran.bst: No hyphenation pattern has been}%
\typeout{** loaded for the language `#1'. Using the pattern for}%
\typeout{** the default language instead.}%
\else
\language=\csname l@#1\endcsname
\fi
#2}}
\providecommand{\BIBdecl}{\relax}
\BIBdecl

\bibitem{PXQ0VW19}
P.~Peng, C.~Xu, L.~Quinn, H.~Hu, B.~Viswanath, and G.~Wang, ``What happens
  after you leak your password: Understanding credential sharing on phishing
  sites,'' in \emph{AsiaCCS}, 2019, pp. 181--192.

\bibitem{hong2012state}
J.~Hong, ``The state of phishing attacks,'' \emph{Commun. ACM}, vol.~55, no.~1,
  pp. 74--81, 2012.

\bibitem{Mathews17}
(2017) Phishing scams cost american businesses half a billion dollars a year.

\bibitem{oest2019phishfarm}
A.~Oest, Y.~Safaei, A.~Doup\'e, G.-J. Ahn, B.~Wardman, and K.~Tyers,
  ``Phishfarm: A scalable framework for measuring the effectiveness of evasion
  techniques against browser phishing blacklists,'' in \emph{IEEE S\&P}, 2019.

\bibitem{teraguchi2004client}
N.~Chou, R.~Ledesma, Y.~Teraguchi, and J.~C. Mitchell, ``Client-side defense
  against web-based identity theft,'' in \emph{{NDSS}}, 2004.

\bibitem{zhang2007cantina}
Y.~Zhang, J.~I. Hong, and L.~F. Cranor, ``Cantina: a content-based approach to
  detecting phishing web sites,'' in \emph{WWW}, 2007.

\bibitem{AfrozG11}
S.~Afroz and R.~Greenstadt, ``Phishzoo: Detecting phishing websites by looking
  at them,'' in \emph{{ICSC}}.

\bibitem{medvet2008visual}
E.~Medvet, E.~Kirda, and C.~Kruegel, ``Visual-similarity-based phishing
  detection,'' in \emph{{SecureComm}}, 2008.

\bibitem{fu2006detecting}
A.~Y. Fu, L.~Wenyin, and X.~Deng, ``Detecting phishing web pages with visual
  similarity assessment based on earth mover's distance ({EMD}),'' \emph{IEEE
  Trans. Dependable Sec. Comput.}, vol.~3, no.~4, pp. 301--311, 2006.

\bibitem{rosiello2007layout}
A.~P. Rosiello, E.~Kirda, F.~Ferrandi \emph{et~al.}, ``A
  layout-similarity-based approach for detecting phishing pages,'' in
  \emph{{SecureComm}}, 2007.

\bibitem{pan2006anomaly}
Y.~Pan and X.~Ding, ``Anomaly based web phishing page detection,'' in
  \emph{{ACSAC}}, 2006, pp. 381--392.

\bibitem{basnet2008detection}
R.~Basnet, S.~Mukkamala, and A.~H. Sung, ``Detection of phishing attacks: {A}
  machine learning approach,'' in \emph{Soft Computing Applications in
  Industry}, 2008, pp. 373--383.

\bibitem{ma2009beyond}
J.~Ma, L.~K. Saul, S.~Savage, and G.~M. Voelker, ``Beyond blacklists: learning
  to detect malicious web sites from suspicious {URLs},'' in \emph{{KDD}},
  2009, pp. 1245--1254.

\bibitem{whittaker2010large}
C.~Whittaker, B.~Ryner, and M.~Nazif, ``Large-scale automatic classification of
  phishing pages,'' in \emph{{NDSS}}, 2010.

\bibitem{XHRC11}
G.~Xiang, J.~I. Hong, C.~P. Ros{\'{e}}, and L.~F. Cranor, ``{CANTINA+:} {A}
  feature-rich machine learning framework for detecting phishing web sites,''
  \emph{{ACM} Trans. Inf. Syst. Secur.}, vol.~14, no.~2, pp. 21:1--21:28, 2011.

\bibitem{corona2017deltaphish}
I.~Corona, B.~Biggio, M.~Contini, L.~Piras, R.~Corda, M.~Mereu, G.~Mureddu,
  D.~Ariu, and F.~Roli, ``Deltaphish: Detecting phishing webpages in
  compromised websites,'' in \emph{{ESORICS}}, 2017.

\bibitem{ubing2019phishing}
A.~A. Ubing, S.~K.~B. Jasmi, A.~Abdullah, N.~Jhanjhi, and M.~Supramaniam,
  ``Phishing website detection: An improved accuracy through feature selection
  and ensemble learning,'' \emph{Int. J. Of Advanced Computer Science and
  Applications}, vol.~10, no.~1, pp. 252--257, 2019.

\bibitem{edge20}
(2020)
  \url{https://docs.microsoft.com/en-us/windows/security/threat-protection/windows-defender-smartscreen/windows-defender-smartscreen-overview}.

\bibitem{SL14}
N.~Srndic and P.~Laskov, ``Practical evasion of a learning-based classifier:
  {A} case study,'' in \emph{{IEEE} S\&P}, 2014, pp. 197--211.

\bibitem{xu2016automatically}
W.~Xu, Y.~Qi, and D.~Evans, ``Automatically evading classifiers: A case study
  on pdf malware classifiers,'' in \emph{NDSS}, 2016, pp. 21--24.

\bibitem{chen2018automated}
S.~Chen, M.~Xue, L.~Fan, S.~Hao, L.~Xu, H.~Zhu, and B.~Li, ``Automated
  poisoning attacks and defenses in malware detection systems: {An} adversarial
  machine learning approach,'' \emph{Computers \& Security}, vol.~73, pp.
  326--344, 2018.

\bibitem{hu2018black}
W.~Hu and Y.~Tan, ``Black-box attacks against {RNN} based malware detection
  algorithms,'' in \emph{AAAI Workshops}, 2018.

\bibitem{liang2016cracking}
B.~Liang, M.~Su, W.~You, W.~Shi, and G.~Yang, ``Cracking classifiers for
  evasion: A case study on the {Google's} phishing pages filter,'' in
  \emph{WWW}, 2016, pp. 345--356.

\bibitem{wroblewski2002general}
G.~Wroblewski, ``General method of program code obfuscation (draft),'' Ph.D.
  dissertation, Wroclaw University of Technology, 2002.

\bibitem{jia2011analysis}
Y.~Jia and M.~Harman, ``An analysis and survey of the development of mutation
  testing,'' \emph{IEEE Trans. Software Eng.}, vol.~37, no.~5, pp. 649--678,
  2011.

\bibitem{godefroid2008automated}
P.~Godefroid, M.~Y. Levin, D.~A. Molnar \emph{et~al.}, ``Automated whitebox
  fuzz testing.'' in \emph{NDSS}, vol.~8, 2008, pp. 151--166.

\bibitem{hazewinkel2001greedy}
M.~Hazewinkel, ``Greedy algorithm,'' \emph{Encyclopedia of Mathematics}, 2001.

\bibitem{MAGSSA17}
S.~Marchal, G.~Armano, T.~Grondahl, K.~Saari, N.~Singh, and N.~Asokan,
  ``Off-the-hook: An efficient and usable client-side phishing prevention
  application,'' \emph{{IEEE} Trans. Computers}, vol.~66, no.~10, pp.
  1717--1733, 2017.

\bibitem{ThomasGMPS11}
K.~Thomas, C.~Grier, J.~Ma, V.~Paxson, and D.~Song, ``Design and evaluation of
  a real-time {URL} spam filtering service,'' in \emph{{IEEE} S\&P}, 2011, pp.
  447--462.

\bibitem{trafficlight}
(2019) \url{https://addons.mozilla.org/en-US/firefox/addon/trafficlight}.

\bibitem{netcraft}
(2019) \url{https://www.netcraft.app/browser-extension}.

\bibitem{360protection}
(2019)
  \url{https://chrome.google.com/webstore/detail/360-internet-protection/glcimepnljoholdmjchkloafkggfoijh?hl=en}.

\bibitem{AF13}
D.~Akhawe and A.~P. Felt, ``Alice in warningland: {A} large-scale field study
  of browser security warning effectiveness,'' in \emph{the 22th {USENIX}
  Security Symposium}, 2013, pp. 257--272.

\bibitem{safebrowsing}
(2019) \url{https://wiki.mozilla.org/Security/Safe_Browsing}.

\bibitem{wb09}
Z.~Wang and A.~C. Bovik, ``Mean squared error: Love it or leave it? a new look
  at signal fidelity measures,'' \emph{IEEE Signal Process. Mag.}, vol.~26,
  no.~1, pp. 98--117, 2009.

\bibitem{chai2014root}
T.~Chai and R.~R. Draxler, ``Root mean square error (rmse) or mean absolute
  error (mae)?--arguments against avoiding rmse in the literature,''
  \emph{Geoscientific model development}, vol.~7, no.~3, pp. 1247--1250, 2014.

\bibitem{LiuDHF06}
W.~Liu, X.~Deng, G.~Huang, and A.~Y. Fu, ``An antiphishing strategy based on
  visual similarity assessment,'' \emph{{IEEE} Internet Computing}, vol.~10,
  no.~2, pp. 58--65, 2006.

\bibitem{Pelican}
(2019) Our dataset. \url{https://sites.google.com/view/xpelican}.

\bibitem{krombholz2015advanced}
K.~Krombholz, H.~Hobel, M.~Huber, and E.~Weippl, ``Advanced social engineering
  attacks,'' \emph{Journal of Information Security and applications}, vol.~22,
  pp. 113--122, 2015.

\bibitem{chen2019gui}
S.~Chen, L.~Fan, C.~Chen, M.~Xue, Y.~Liu, and L.~Xu, ``Gui-squatting {Attack}:
  Automated generation of {Android} phishing apps,'' \emph{IEEE Transactions on
  Dependable and Secure Computing}, 2019.

\bibitem{most}
(2017) \url{http://www.phishing.org/phishing-techniques}.

\bibitem{kirda2006protecting}
E.~Kirda and C.~Kruegel, ``Protecting users against phishing attacks,''
  \emph{The Computer Journal}, vol.~49, no.~5, pp. 554--561, 2006.

\bibitem{gupta2018defending}
B.~Gupta, N.~A. Arachchilage, and K.~E. Psannis, ``Defending against phishing
  attacks: taxonomy of methods, current issues and future directions,''
  \emph{Telecommunication Systems}, vol.~67, no.~2, pp. 247--267, 2018.

\bibitem{tang2019large}
C.~Tang, S.~Chen, L.~Fan, L.~Xu, Y.~Liu, Z.~Tang, and L.~Dou, ``A large-scale
  empirical study on industrial fake apps,'' in \emph{Proceedings of the 41st
  International Conference on Software Engineering: Software Engineering in
  Practice}.\hskip 1em plus 0.5em minus 0.4em\relax IEEE Press, 2019, pp.
  183--192.

\bibitem{VMA16}
G.~Varshney, M.~Misra, and P.~K. Atrey, ``A survey and classification of web
  phishing detection schemes,'' \emph{Security and Communication Networks},
  vol.~9, no.~18, pp. 6266--6284, 2016.

\bibitem{ludl2007effectiveness}
C.~Ludl, S.~McAllister, E.~Kirda, and C.~Kruegel, ``On the effectiveness of
  techniques to detect phishing sites,'' in \emph{DIMVA}, 2007, pp. 20--39.

\bibitem{chikofsky1990reverse}
E.~J. Chikofsky and J.~H. Cross, ``Reverse engineering and design recovery: A
  taxonomy,'' \emph{IEEE software}, vol.~7, no.~1, pp. 13--17, 1990.

\bibitem{wang2005break}
X.~Wang and H.~Yu, ``How to break {MD5} and other hash functions,'' in
  \emph{EUROCRYPT}, 2005, pp. 19--35.

\bibitem{PMGJCS17}
N.~Papernot, P.~D. McDaniel, I.~J. Goodfellow, S.~Jha, Z.~B. Celik, and
  A.~Swami, ``Practical black-box attacks against machine learning,'' in
  \emph{AsiaCCS}, 2017, pp. 506--519.

\bibitem{BHLS18}
A.~N. Bhagoji, W.~He, B.~Li, and D.~Song, ``Practical black-box attacks on deep
  neural networks using efficient query mechanisms,'' in \emph{{ECCV}}, 2018,
  pp. 158--174.

\bibitem{phishtank}
(2019) \url{https://www.phishtank.com}.

\bibitem{phishnet}
(2019) \url{http://203.80.17.57/phishnet/individual}.

\bibitem{HAJSS14}
W.~He, D.~Akhawe, S.~Jain, E.~Shi, and D.~X. Song, ``Shadowcrypt: Encrypted web
  applications for everyone,'' in \emph{CCS}, 2014.

\bibitem{safebrowsingAPI}
(2019) \url{https://developers.google.com/safe-browsing/v4/}.

\bibitem{depottools19}
(2019) \url{https://www.chromium.org/developers/how-tos/depottools}.

\bibitem{ninja}
(2019) \url{https://github.com/ninja-build/ninja}.

\bibitem{maiorca2013looking}
D.~Maiorca, I.~Corona, and G.~Giacinto, ``Looking at the bag is not enough to
  find the bomb: an evasion of structural methods for malicious pdf files
  detection,'' in \emph{AsiaCCS}, 2013, pp. 119--130.

\bibitem{grosse2017adversarial}
K.~Grosse, N.~Papernot, P.~Manoharan, M.~Backes, and P.~McDaniel, ``Adversarial
  examples for malware detection,'' in \emph{{ESORICS}}, 2017.

\bibitem{demontis2017yes}
A.~Demontis, M.~Melis, B.~Biggio, D.~Maiorca, D.~Arp, K.~Rieck, I.~Corona,
  G.~Giacinto, and F.~Roli, ``Yes, machine learning can be more secure! a case
  study on android malware detection,'' \emph{IEEE Trans. Dependable Sec.
  Comput.}, 2017.

\bibitem{xu2014evasion}
L.~Xu, Z.~Zhan, S.~Xu, and K.~Ye, ``An evasion and counter-evasion study in
  malicious websites detection,'' in \emph{CNS}, 2014, pp. 265--273.

\bibitem{lowd2005good}
D.~Lowd and C.~Meek, ``Good word attacks on statistical spam filters.'' in
  \emph{CEAS}, vol. 2005, 2005.

\bibitem{chen2019real}
G.~Chen, S.~Chen, L.~Fan, X.~Du, Z.~Zhao, F.~Song, and Y.~Liu, ``Who is real
  {Bob}? adversarial attacks on speaker recognition systems,'' \emph{arXiv
  preprint arXiv:1911.01840}, 2019.

\bibitem{SBR18}
M.~Sharif, L.~Bauer, and M.~K. Reiter, ``On the suitability of lp-norms for
  creating and preventing adversarial examples,'' in \emph{CVPR Workshops},
  2018, pp. 1605--1613.

\bibitem{SBBR19}
M.~Sharif, S.~Bhagavatula, L.~Bauer, and M.~K. Reiter, ``A general framework
  for adversarial examples with objectives,'' \emph{{ACM} Trans. Priv. Secur.},
  vol.~22, no.~3, pp. 16:1--16:30, 2019.

\bibitem{carlini2017adversarial}
N.~Carlini and D.~Wagner, ``Adversarial examples are not easily detected:
  Bypassing ten detection methods,'' in \emph{AISec@CCS}, 2017, pp. 3--14.

\bibitem{biggio2012poisoning}
B.~Biggio, B.~Nelson, and P.~Laskov, ``Poisoning attacks against support vector
  machines,'' in \emph{ICML}, 2012.

\bibitem{nelson2008exploiting}
B.~Nelson, M.~Barreno, F.~J. Chi, A.~D. Joseph, B.~I.~P. Rubinstein, U.~Saini,
  C.~A. Sutton, J.~D. Tygar, and K.~Xia, ``Exploiting machine learning to
  subvert your spam filter,'' in \emph{{LEET}}, 2008.

\bibitem{biggio2014security}
B.~Biggio, G.~Fumera, and F.~Roli, ``Security evaluation of pattern classifiers
  under attack,'' \emph{IEEE Trans. Knowl. Data Eng.}, vol.~26, no.~4, pp.
  984--996, 2014.

\bibitem{saeys2008robust}
Y.~Saeys, T.~Abeel, and Y.~Van~de Peer, ``Robust feature selection using
  ensemble feature selection techniques,'' in \emph{Int. J. Machine Learning \&
  Cybernetics}, 2008, pp. 313--325.

\bibitem{biggio2010multiple}
B.~Biggio, G.~Fumera, and F.~Roli, ``Multiple classifier systems for robust
  classifier design in adversarial environments,'' \emph{Int. J. of Machine
  Learning \& Cybernetics}, vol.~1, no. 1-4, pp. 27--41, 2010.

\bibitem{smutz2016tree}
C.~Smutz and A.~Stavrou, ``When a tree falls: Using diversity in ensemble
  classifiers to identify evasion in malware detectors.'' in \emph{NDSS}, 2016.

\bibitem{goodfellow2014explaining}
I.~J. Goodfellow, J.~Shlens, and C.~Szegedy, ``Explaining and harnessing
  adversarial examples,'' in \emph{ICLR (Poster)}, 2015.

\bibitem{zhang2016adversarial}
F.~Zhang, P.~P. Chan, B.~Biggio, D.~S. Yeung, and F.~Roli, ``Adversarial
  feature selection against evasion attacks,'' \emph{IEEE Trans. Cybernetics},
  vol.~46, no.~3, pp. 766--777, 2016.

\bibitem{chen2016towards}
S.~Chen, M.~Xue, and L.~Xu, ``Towards adversarial detection of mobile
  malware,'' in \emph{Proceedings of the 22nd Annual International Conference
  on Mobile Computing and Networking}.\hskip 1em plus 0.5em minus 0.4em\relax
  ACM, 2016, pp. 415--416.

\bibitem{liang2017detecting}
B.~Liang, H.~Li, M.~Su, X.~Li, W.~Shi, and X.~Wang, ``Detecting adversarial
  examples in deep networks with adaptive noise reduction,'' \emph{IEEE Trans.
  Dependable Sec. Comput.}, 2020.

\bibitem{xu2017feature}
W.~Xu, D.~Evans, and Y.~Qi, ``Feature squeezing: Detecting adversarial examples
  in deep neural networks,'' in \emph{{NDSS}}, 2018.

\bibitem{SBBR16}
M.~Sharif, S.~Bhagavatula, L.~Bauer, and M.~K. Reiter, ``Accessorize to a
  crime: Real and stealthy attacks on state-of-the-art face recognition,'' in
  \emph{CCS}, 2016, pp. 1528--1540.

\bibitem{ASCZHS19}
M.~Alzantot, Y.~Sharma, S.~Chakraborty, H.~Zhang, C.~Hsieh, and M.~B.
  Srivastava, ``Genattack: practical black-box attacks with gradient-free
  optimization,'' in \emph{GECCO}, 2019, pp. 1111--1119.

\end{thebibliography}

\begin{IEEEbiography}[{\includegraphics[width=1in,height=1.25in,clip,keepaspectratio]{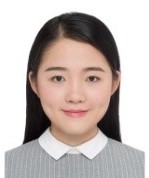}}]{Yusi Lei} 
received the B.S. degree and Master degree in software engineering from East China Normal University, China, in 2017 and 2020. 
She is currently a research assistant in ShanghaiTech University. Her research interests include web security and machine learning.
\end{IEEEbiography}

\begin{IEEEbiography}[{\includegraphics[width=1in,height=1.25in,clip,keepaspectratio]{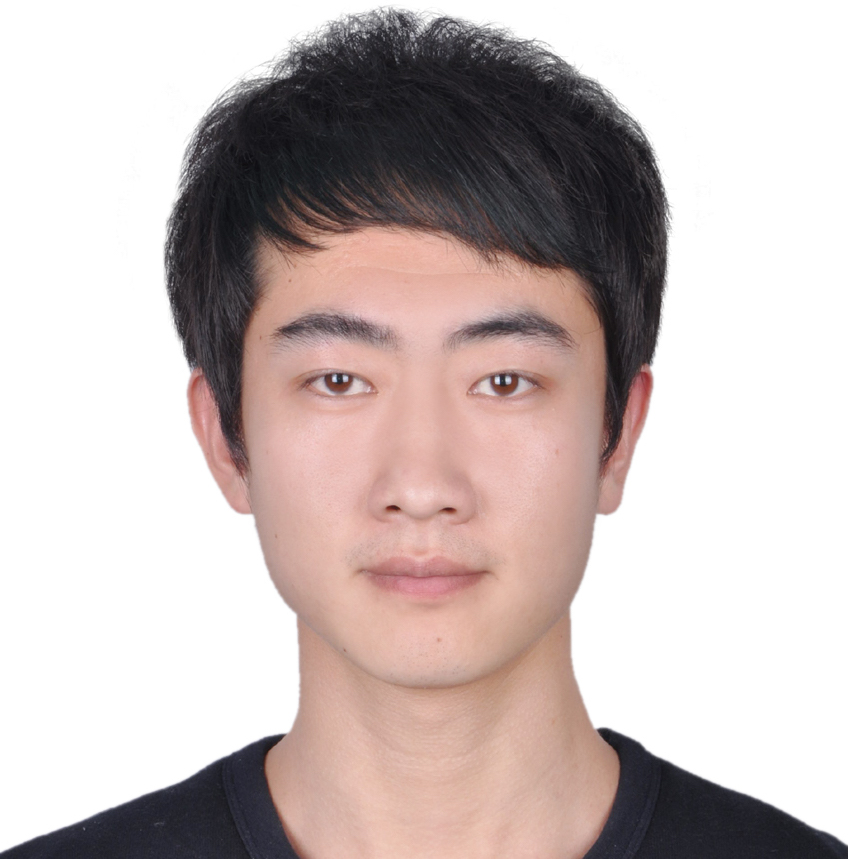}}]
	{Sen Chen} received his Ph.D. degree in computer science from School of Computer Science and Software Engineering, East China Normal University, Shanghai, China, in 2019.
	Currently, he is a Research Fellow in School of Computer Science and Engineering, Nanyang Technological University, Singapore.
	Previously, he was a visiting scholar of Cybersecurity Lab, SCSE, NTU from October 2016 to June 2019.
	His research focuses on software engineering, security, and data-driven analytics.
	He has published broadly in top-tier security and software engineering venues, including Oakland, USENIX Security, ICSE, ESEC/FSE, ASE, and TDSC.
	He got an ACM SIGSOFT Distinguished Paper Award at ICSE 2018.
	More information is available on {https://sen-chen.github.io/}
\end{IEEEbiography}

\begin{IEEEbiography}[{\includegraphics[width=1in,height=1.25in,clip,keepaspectratio]{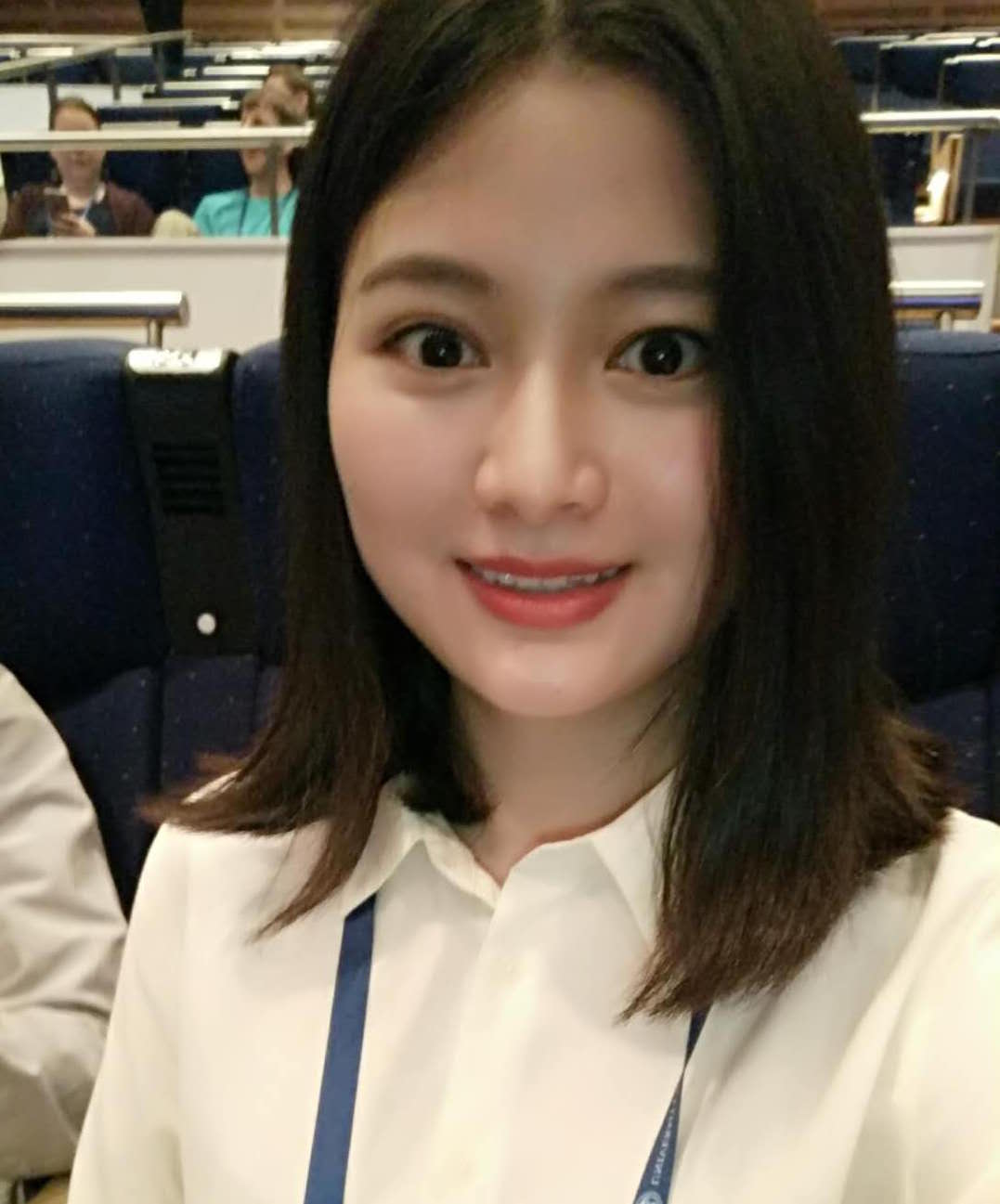}}]
	{Lingling Fan} received her Ph.D and B.S. degrees in computer science from East China Normal University, Shanghai, China in June 2019 and June 2014, respectively, and now she is a postdoctoral Research Fellow in School of Computer Science and Engineering, Nanyang Technological University, Singapore. Her research focuses on program analysis and testing, Android application security analysis and testing, and model checking. She got an ACM SIGSOFT Distinguished Paper Award at ICSE 2018 as the first author. More information is available on https://lingling-fan.github.io/
\end{IEEEbiography}

\begin{IEEEbiography}[{\includegraphics[width=1in,height=1.25in,clip,keepaspectratio]{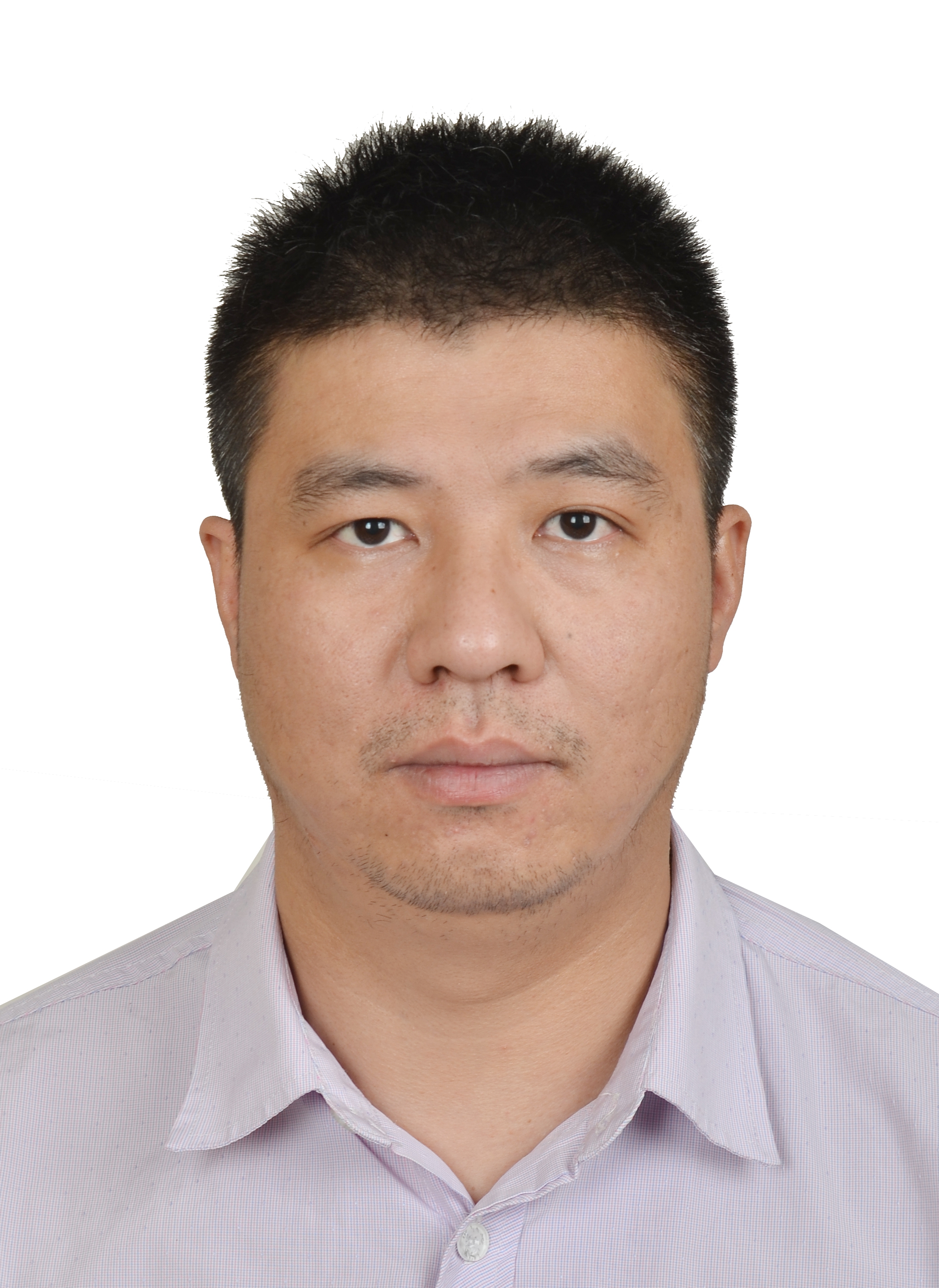}}]{Fu Song}
received the B.S. degree from Ningbo
University, Ningbo, China, in 2006, the M.S. degree from East China Normal University, Shanghai,
China, in 2009, and the Ph.D. degree in computer science from University Paris-Diderot,
Paris, France, in 2013.

From 2013 to 2016, he was a Lecturer and Associate Research Professor at East China Normal University.
Since August 2016, he is an Assistant Professor with ShanghaiTech University, Shanghai, China. His research
interests include formal methods and computer security, especially about automata, logic, model checking, and program analysis.
Dr. Song was a recipient of EASST best paper award at ETAPS 2012.
\end{IEEEbiography}

\begin{IEEEbiography}[{\includegraphics[width=1in,height=1.25in,clip,keepaspectratio]{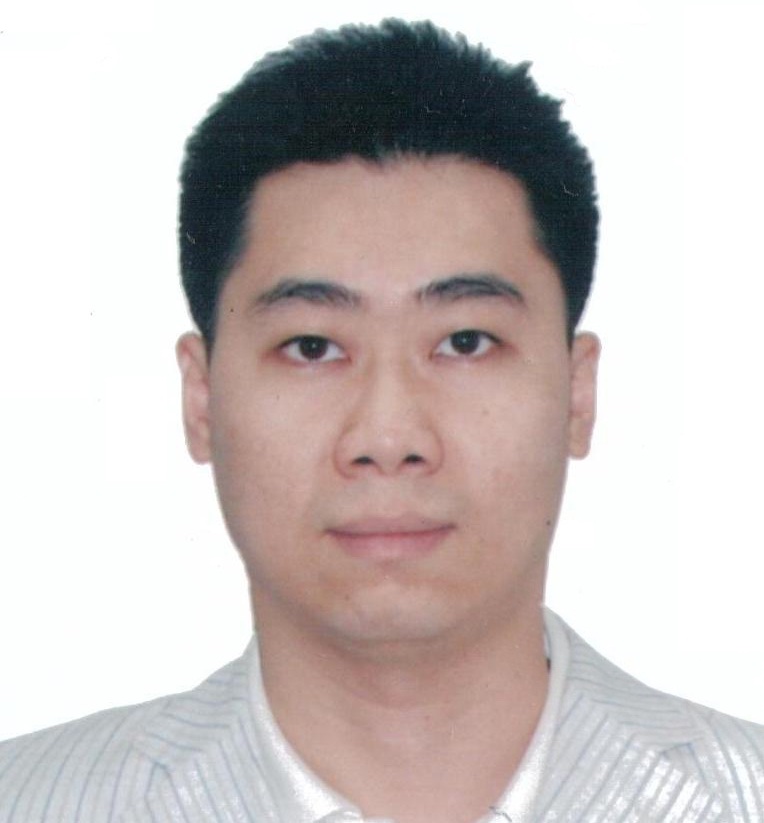}}]{Liu Yang}
	graduated in 2005 with a Bachelor of Computing (Honours) in the National University of Singapore (NUS). In 2010, he obtained his PhD and started his post doctoral work in NUS, MIT and SUTD. In 2011, Dr. Liu is awarded the Temasek Research Fellowship at NUS to be the Principal Investigator in the area of Cyber Security. In 2012 fall, he joined Nanyang Technological University (NTU) as a Nanyang Assistant Professor. He is currently a full professor and the director of the cybersecurity lab in NTU.

He specializes in software verification, security and software engineering. His research has bridged the gap between the theory and practical usage of formal methods and program analysis to evaluate the design and implementation of software for high assurance and security. His work led to the development of a state-of-the-art model checker, Process Analysis Toolkit (PAT). By now, he has more than 300 publications and 6 best paper awards in top tier conferences and journals. With more than 20 million Singapore dollar funding support, he is leading a large research team working on the state-of-the-art software engineering and cybersecurity problems.
\end{IEEEbiography}

\end{document}